\documentclass[aip,graphicx]{revtex4-1}
\usepackage{graphicx}
\usepackage{amsmath}

\draft % marks overfull lines with a black rule on the right

\begin{document}

% Use the \preprint command to place your local institutional report number 
% on the title page in preprint mode.
% Multiple \preprint commands are allowed.
%\preprint{}

\title{Numerical dissipation induced by the low-pass filtering in nonlinear gyrokinetic simulations} %Title of paper

% repeat the \author .. \affiliation  etc. as needed
% \email, \thanks, \homepage, \altaffiliation all apply to the current author.
% Explanatory text should go in the []'s, 
% actual e-mail address or url should go in the {}'s for \email and \homepage.
% Please use the appropriate macro for the type of information

% \affiliation command applies to all authors since the last \affiliation command. 
% The \affiliation command should follow the other information.

\author{Zihao Wang}
%\email[]{Your e-mail address}
%\homepage[]{Your web page}
%\thanks{}
%\altaffiliation{}
\affiliation{Department of Engineering and Applied Physics, University of Science and Technology of China, Hefei, 230026, China}

\author{Shaojie Wang}
\email{wangsj@ustc.edu.cn}
%\homepage[]{Your web page}
%\thanks{}
%\altaffiliation{}
\affiliation{Department of Engineering and Applied Physics, University of Science and Technology of China, Hefei, 230026, China}

% Collaboration name, if desired (requires use of superscriptaddress option in \documentclass). 
% \noaffiliation is required (may also be used with the \author command).
%\collaboration{}
%\noaffiliation

\date{\today}

\begin{abstract}
% insert abstract here
  De-aliasing is an essential procedure for eliminating the aliasing error in nonlinear simulations, such as nonlinear gyrokinetic turbulence simulations.
  An ideal approach to de-aliasing in the periodic dimension is Fourier truncation.
  Finite difference low-pass filtering applied in the non-periodic direction strongly dampens aliasing modes. At the same time, it induces numerical dissipation in the region of the physically realistic solution.
  It is shown analytically that the long-wave dissipation coefficient is proportional to the ($N_{p}-3$) power of the wavenumber under desirable constraints satisfying the highest order accuracy, where $N_{p}$ is the number of filter points.
  Numerical results after applying the optimised low-pass filtering to the nonlinear gyrokinetic turbulence simulation suggest that the nine-point format preserves intact mesoscopic zonal structures in Tokamak plasmas, and is therefore suitable for long-time nonlinear turbulence simulations.
  
\end{abstract}

% insert suggested PACS numbers in braces on next line

\maketitle %\maketitle must follow title, authors, abstract and \pacs

% Body of paper goes here. Use proper sectioning commands. 
% References should be done using the \cite, \ref, and \label commands
\section{Introduction}

  The numerical solution of nonlinear complex dynamical systems with discrete grids results in the emergence of the so-called aliasing mode~\cite{tajima2018computational}, which is rooted in the incompatibility between discrete Eulerian grids and continuous Lagrangian coordinates.
  The nonlinear product of perturbations in coordinate space becomes convolution in wave vector space after Fourier transform, and aliasing occurs when the convolution wavenumber exceeds the cutoff wavenumber (equal to half the number of grid points in one spatial dimension).
  In addition, all the discrete derivative operators cannot accurately handle the short-wave components.
  Aliasing errors cause incorrect computation results and even numerical instability in high-wavenumber (high-$k$) modes.
  In nonlinear fluid dynamics simulations for solving the nonlinear Navier-Stokes equation represented by the large eddy simulation (LES)~\cite{deardorff1970numerical,schumann1975subgrid}, the turbulence scale is computed only up to the cutoff wavenumber, whereas aliasing errors significantly interfere with the numerical results of small-scale eddies.
  In nonlinear gyrokinetic simulations for solving the Vlasov-Poisson system, the spectral particle-in-cell approach~\cite{birdsall2018plasma} localizes the coupling between grids by constructing smooth and continuous particle shape functions, but aliasing errors destroy the energy conservation property and lead to self-heating of the girds~\cite{langdon1970effects,godfrey1974numerical,xu2013numerical,huang2016finite}.
  Nonlinear mode couplings in the Eulerian or semi-Lagrangian approaches cause numerical oscillation and divergence as well~\cite{maeyama2013numerical,xu2022gyrokinetic}.

\begin{figure}[htp]
      \centering
      \includegraphics[width=0.9\textwidth]{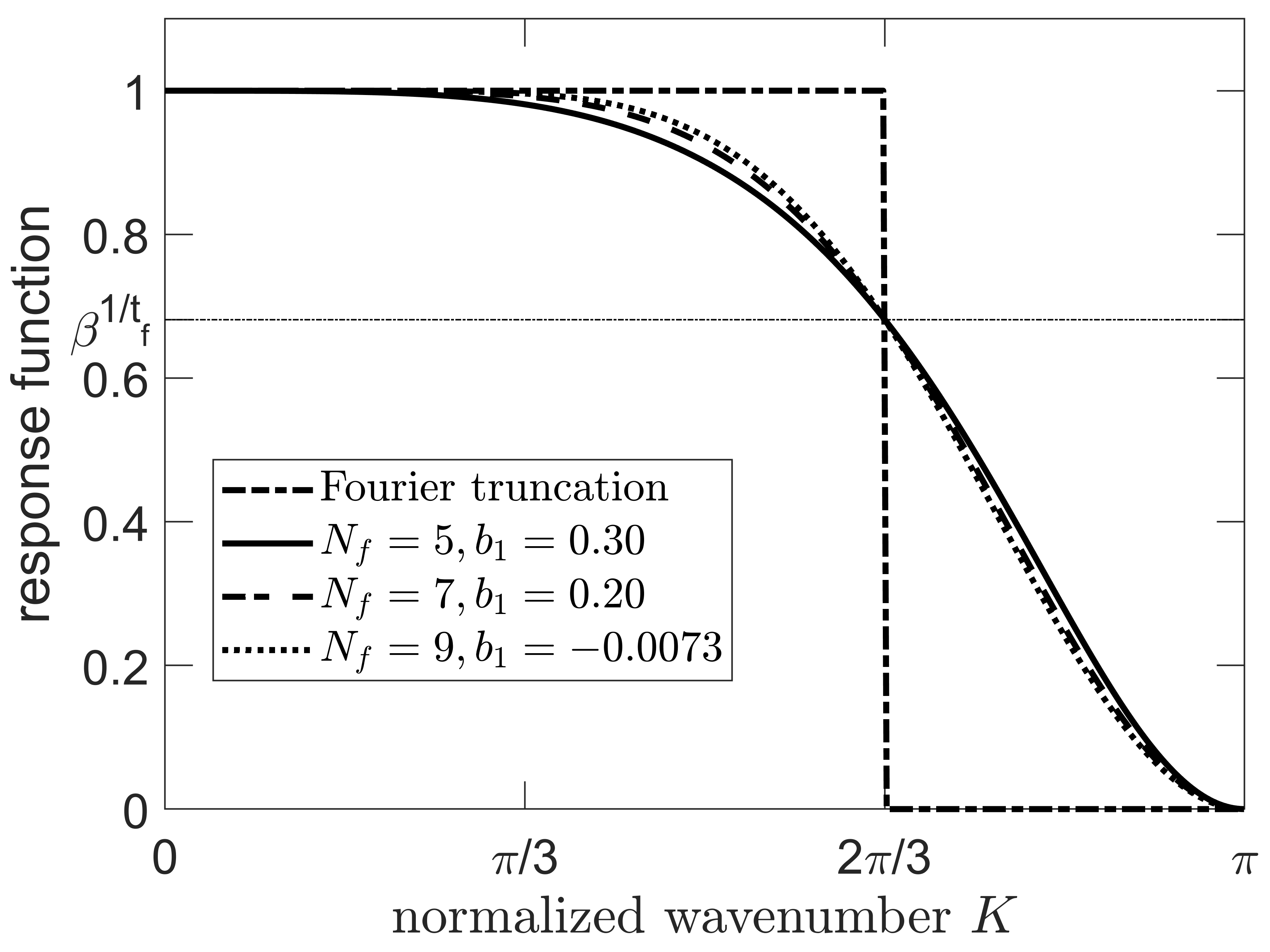}
      \caption{Response functions $G(K)$ of the $3\Delta x$ rule filter (dash-dotted) and the symmetric Pade-type discrete filters with five (solid), seven (dashed), nine (dotted) filter points corresponding respectively to cases 1,2,3 given in Table I, which share the same value, $\beta=0.1$, at $K_{0} = 2\mathrm{\pi}/3$; the parameter in Eq.~\ref{final_constraint} is $N_{f}=6$.}
      \label{fig:response}
  \end{figure}

  The process of eliminating aliasing errors is often referred to as de-aliasing~\cite{phillips1959example}.
  Early developments for removing pseudo-spectral convolution involved the use of the phase-shift method~\cite{patterson1971spectral} or the zero-padding strategy~\cite{orszag1971elimination}.
  The latter de-aliases by zeroing out the top one-third of components contaminated by aliasing modes, also known as the $2/3$ rule or the $3\Delta x$ rule;
  nevertheless, these ideal approaches are only applicable in the periodic dimension where the Fourier series is well-defined, and the abrupt truncation of the high-$k$ spectrum using Fourier filtering in a non-periodic direction induces the well-known Gibbs phenomenon.
  An alternative strategy is to construct numerical filters in finite difference format to reduce aliasing errors (hereafter called low-pass filters).
  Since the Schumann filter was first proposed~\cite{shuman1957numerical}, it has been recognized that low-pass filtering must be used to ensure numerical stability and minimize the impact of filters on the resolution of physical results.
  This is because the response function of low-pass filters is not exactly the same as the ideal truncation function in the region of relatively long wavelengths, which can be seen from Fig.~\ref{fig:response}.
  This introduces an additional artificial dissipation term, which, if not evaluated, will affect the reliability of the numerical results.
  In particular, it is of interest to develop high-precision filters with minimal dissipation in nonlinear turbulence simulations in the case of long-time nonlinear evolution~\cite{strugarek2013unraveling,imadera2022itb,wang2024self}, integration~\cite{tam1995computational}, or short-scale shock calculations~\cite{pirozzoli2006spectral,shu2020essentially}.
  Applicable in the non-periodic direction, the implicit compact filters proposed by Lele take into account both the higher-order accuracy and bandwidth characteristics, which weakly dissipate all longer-wave modes under a specified criterion~\cite{lele1992compact}.
  A general set of rules has also been developed for constructing discrete filters in turbulent inhomogeneous flows and complex geometries~\cite{vasilyev1998general}.

  These low-pass filters are widely used in LES, but in massively parallel gyrokinetic simulations of Tokamak plasma, the evaluation of system reliability by low-pass filters has not been carried out~\cite{lin1998turbulent,chen2003deltaf,idomura2003global,jolliet2007global,heikkinen2008full,wang2010nonlinear}.
  Moreover, this assessment is necessary, not only for the well-known consideration that low-pass filtering numerically dissipates fast-developing turbulence but, more importantly, it furthermore weakens zonal structures, which have longer characteristic lengths compared to turbulence.
  Symmetrical zonal structures~\cite{chen2007nonlinear,falessi2019transport} in toroidal fusion plasmas, such as temperature profiles, density profiles, and zonal flows~\cite{hasegawa1987self,diamond2005zonal}, serve as the long-lived and slowly time-varying neighboring nonlinear equilibria~\cite{chen2007nonlinear,wang2024self} relative to microscopic perturbations, effectively redefining the characteristic spatiotemporal scales of the system and determining the statistical properties of critical transport events, such as intermittent outbursts, avalanches, and non-local behavior.
  Therefore, the long lifetime of radial zonal structures means that they will continue to be artificially dissipated and the mesoscopic structures destroyed by the low-pass filter.
  For these considerations, low-pass filters should have to correctly resolve multiple characteristic scales of Tokamak plasma turbulent transport.
  %Furthermore, the toroidal and poloidal symmetric zonal flows, as a long-time stable mesoscale structure, are widely recognized as a crucial role in regulating turbulence level, reducing anomalous transport, and improving confinement. Therefore, multiples successive low-pass filtering prohibits disrupting the mesoscale structure of zonal flows in the aperiodic radial direction.
  
  With such a goal in mind, in this paper, we derive that the diffusion coefficient $\chi(K)$ introduced by low-pass filters in the region of the physically realistic solution is proportional to $K^{N_{p}-3}$, where $K$ and $N_{p}$ are the normalized wavenumbers and the number of filter points, respectively.
  This suggests that the finite difference scheme with more points better preserves both the microscale turbulence and meso- to macro-scale zonal structures.
  We apply the optimized low-pass filtering to the non-periodic radial direction of the massively parallel gyrokinetic code, NLT, and numerically simulate the relaxation process with ion-temperature-gradient (ITG) turbulence in Tokamak plasmas.
  The results show that the different formats of low-pass filtering barely affect the linear growth rate of the ITG turbulence, but the nonlinear evolutionary behavior and the level of anomalous transport of the ITG turbulence are not the same.
  This is because low-pass filtering with fewer points significantly destroys the mesoscale zonal radial electric field and temperature profile in long-time simulation and indirectly alters the turbulence instability at the neighboring nonlinear equilibrium.
  The nine-point format preserves intact mesoscopic zonal structures in Tokamak plasmas, and is therefore suitable for long-time nonlinear turbulence simulations.

  This work is organized as follows: in Sec.~\ref{Equivalent}, we present a general set of difference schemes for constructing discrete filters and quantitatively give the equivalent diffusion coefficient introduced by low-pass filtering.
  In Sec.~\ref{Simulations}, We compare the nonlinear evolution of turbulence and zonal structures in different filter formats after application to the gyrokinetic code.
  Final conclusions and discussions are given in Sec.~\ref{Conclution}.
  
\section{Equivalent diffusion coefficient of low-pass filters} \label{Equivalent}
  Consider a one-dimensional continuously integrable function $\psi(x)$ defined in the domain $[a,b]$. $\{\psi_{j}\}$ corresponds to the values of $\psi(x_{j})$ at location $x_{j}=a+(j-1)\Delta, (j = 1,2,...,N_{x})$, where $\Delta = (b-a)/N_{x}$ is the uniform grid spacing and $N_{x}$ is the number of sampling grid points.
  If the sampling grid is non-uniform, consider mapping the non-uniform computational grid on $\bar{x} \in [\alpha,\beta]$ into the uniform grid on $x \in [a,b]$ by using a monotonic differentiable function $x = \mathcal{T}(\bar{x})$; the original function under non-uniform discretization satisfies $\psi(x)=\psi[\mathcal{T}(\bar{x})]=\bar{\psi}(\bar{x})$, where the non-uniform grid spacing is given by $\bar{\Delta}=\Delta/\mathcal{T}'(\bar{x})$.
  In the non-uniform scenario, different schemes of low-pass filters can be applied in different intervals within the domain.
  Therefore, it is without loss of generality to consider finite differences with uniform sampling.

  Given the values of a function $\{\psi_{i}\}$ on a set of nodes the finite difference approximation to the low-pass filtering of the function is expressed as a linear combination of given function values.
  The generalized form of low-pass filters with compact finite difference schemes in the interior of the domain (boundary to be discussed later) can be written as
  \begin{equation} \label{GF}
      \sum_{m=-m_{l}}^{+m_{u}} b_{m}\phi_{j+m} = \sum_{n=-n_{l}}^{+n_{u}} c_{n}\psi_{j+n}.
  \end{equation}
  Here, $\{\phi_{j}\}$ represent the filtered values at grids $\{x_{j}\}$; $m_{l}, m_{u} (n_{l}, n_{u})$ mark the lower and upper bounds of compact (finite difference) scheme, respectively; $b_{m}$ and $c_{n}$ are the coefficients of compact and difference, respectively.
  Define the number of compact points $M_{p}=m_{l}+m_{u}+1$ and the number of finite difference points $N_{p}=n_{l}+n_{u}+1$.

  \subsection{Pade-type low-pass filters}
  In this subsection, we show how to determine the format and coefficients of a low-pass filter. In a one-shot low-pass filter operation, the filtered function value $\phi_{j}$ depends on only a few nearby grids, whereas in the spectral strategy, the value of $\phi_{j}$ depends on all the nodal values.
  The compact scheme mimic this global dependence.
  In order to introduce no or minimal dispersion, the filter operator should be symmetric, i.e. the following relation should be satisfied:
  \begin{equation}
  \begin{aligned}
      m_{l} = m_{u} = (M_{p}-1)/2,& \quad n_{l} = n_{u} = (N_{p}-1)/2 = N_{ph}; \\
      b_{m}=b_{-m},&\quad c_{n} = c_{-n}.
  \end{aligned}
  \end{equation}
  In this way, the low-pass filtering only modifies the amplitude of a given wavelength without distorting its phase.
  One useful example of the compact finite difference algorithm is to use Pade-type filters, which simplifies Eq.~\eqref{GF} by further setting $M_{p}=3$ and $b_{0}=1.0$.
  Eq.~\eqref{GF} can then be rewritten as
  \begin{equation} \label{Pade-type}
      \phi_{j} + b_{1}(\phi_{j+1}+\phi_{j-1}) = c_{0}\psi_{j} + \sum_{n=1}^{N_{ph}} c_{n}(\psi_{j+n}+\psi_{j-n}).
  \end{equation}

  Although the application of low-pass filters is preferable in the non-periodic direction, it is nevertheless necessary to evaluate such operations in wavenumber (Fourier) space.
  From a physical point of view, it is clearly more attractive for the finite difference format  to approximate the Fourier truncation filter, which ensures that aliasing errors are eliminated without impacting on the physically realistic solution.
  The response function is defined as the ratio of amplitude for a given wavenumber before and after filtering, i.e. $G(k) = \phi_{k}/\psi_{k}$.
  Here, $\phi_{k}$ and $\psi_{k}$ are the discrete Fourier transforms of $\phi(x)$ and $\psi(x)$, respectively.
  It is anticipated that the compact schemes provide better control over the shape of response function in wavenumber space and are considerably better approximations of sharp cutoff filters.
  The Fourier transform $G(K)$ associated with Pade-type filters is given by
  \begin{equation} \label{response}
      G(K) = \left(c_{0} + \sum_{n=1}^{N_{ph}} 2c_{n} \mathrm{cos}(nK) \right) \bigg/ \left(1+2b_{1}\mathrm{cos}K\right).
  \end{equation}
  Here, the normalized wavenumbers $K \equiv k\Delta \in [0, \mathrm{\pi}]$.
  The filtering application discussed here naturally is the removal of small scales which are shorter in wavelength than the aliasing modes.
  However, in contrast to Ref.~\onlinecite{lele1992compact} where the desirable characteristics of the response function are more efficiently optimised (the truncation function) without insisting on the highest formal accuracy, this paper consistently requires the response function to have minimal dissipation in the long wavelength region, i.e., in mesoscopic and macroscopic scales.
  This insistence is based on the fact that zonal structures on the mesoscale in Tokamak plasmas play a crucial role in turbulence evolution and anomalous transport~\cite{chen2007nonlinear,wang2024self}.

  The relations between the $N_{ph}+2$ coefficients ($b_{1},c_{0},c_{1},...,c_{N_{ph}}$) are derived by matching the Taylor series coefficients of highest order and additional given constraints.
  Specifically, for the centre-symmetric low-pass filtering, firstly, the constraint of the highest order accuracy ($\mathcal{O}(K^{N_{p}-1})$) in the long-wave region consumes $N_{ph}$ degrees of freedom; secondly, the desirable constraint on a filter is that its Fourier transform be zero at the Nyquist frequency, which guarantees the complete elimination of the most dangerous aliasing mode.
  Thus these $N_{ph}+1$ constraints allow to express ${c_{n}}, (n = 0,1,...,N_{ph})$ as linear functions of $b_{1}$.
  The detailed derivations are described in appendix~\ref{derivation}.
  %Here as an example of the finite difference schemes with $N_{p}=9$, relations between the coefficients are
  %\begin{align}
  %  c_{0} = \frac{93+70b_{1}}{128}, \quad c_{1} = \frac{7+18b_{1}}{32}, \quad 
  %  c_{2} = \frac{-7+14b_{1}}{64}, \quad c_{3} = \frac{1-2b_{1}}{32}, \quad
  %  c_{4} = \frac{-1+2b_{1}}{256}.
  %\end{align}
  
  The last compact parameter ($b_{1}$) is determined by an artificial condition.
  A natural constraint for long-time nonlinear turbulence simulations with variable time advance steps is that the value of the artificial dissipation introduced by low-pass filtering for a given wavelength does not change with the simulation time step.
  For example, the amplitude of a mode with its wavelength three times the grid width, i.e. $\lambda=3\Delta$, decreases to $\beta$ times the original amplitude per unit filtering time, that is
  \begin{equation} \label{final_constraint}
      G(K_{0})^{N_{f}} = \beta.
  \end{equation}
  Here, the wavelength limit at which aliasing occurs $K_{0}=2\mathrm{\pi}/3$, $N_{f}$ is the times that a simulation advances per unit filtering time.
  Examples of response functions for low-pass filters with highest order accuracy and $N_{p}=5,7,9$ are presented in Fig~\ref{fig:response}.
  
  \subsection{Equivalent diffusion coefficient}
  In this subsection, we quantitatively derive the diffusion coefficient introduced by low-pass filtering.
  Consider a standard one-dimensional diffusion equation,
  \begin{equation} \label{diffusion}
      \frac{\partial}{\partial t}\psi(x,t) = \chi \frac{\partial^{2}}{\partial x^{2}}\psi(x,t),
  \end{equation}
  where the diffusion coefficient $\chi$ is usually a constant for slowly varying fields in physics.
  The solution of Eq.~\eqref{diffusion} obtained by Fourier transform, $\psi(x,t)=\psi_{k}(t) \mathrm{e}^{-\mathrm{i}kx}$, is formally written as
  \begin{equation} \label{psik1}
      \psi_{k}(t) = \psi_{k}(t=0)\cdot \mathrm{e}^{-\chi \cdot k^{2}t }.
  \end{equation}
  Low-pass filters numerically dissipates long waves in the non-aliased region.
  After operating successive low-pass filters on $\psi$ with $N_{f}$ times, its amplitude becomes
  \begin{equation} \label{psik2}
      \phi_{k} = G(k)^{N_{f}} \cdot \psi_{k}.
  \end{equation}
  The relationship between the diffusion coefficient and the response function can be obtained by associating Eqs.~(\ref{psik1}-\ref{psik2}), that is
  \begin{equation} \label{chi}
      \chi(k) = -\frac{N_{f}\ln G(k)}{k^{2}t}.
  \end{equation}
  Here, $\chi$ is related to the wavenumber of modes; in general, the artificial diffusion coefficient introduced by low-pass filtering is larger for short wavelengths.
  Due to the constraints on the low-pass filter coefficients that require the highest order accuracy to be satisfied in the long wave region, the response function is formally written in the long wave region as 
  \begin{equation} \label{response_approximate}
      G = 1 - \epsilon \cdot K^{N_{p}-1} + \mathcal{o}(K^{N_{p}-1}) \approx 1 - \epsilon\cdot (k\Delta)^{N_{p}-1}.
  \end{equation}
  The detailed form of $\epsilon$ is given in appendix~\ref{derivation}.
  Hence, by using the approximation $\mathrm{ln}G(k) \approx \epsilon\cdot (k\Delta)^{N_{p}-1}$, the equivalent diffusion coefficient obtained from Eq.~\eqref{chi} is
  \begin{equation} \label{chik}
      \chi(k) \approx \epsilon N_{f,0}\cdot \Delta^{N_{p}-1}\cdot k^{N_{p}-3} \propto k^{N_{p}-3}.
  \end{equation}
  Here, $N_{f,0}=N_{f}/t$ and the scaling factor is related to the format of the low-pass filter and the parameter settings of simulations.
  Importantly, the artificial diffusivity introduced by low-pass filtering is proportional to the ($N_{p}-3$) power of the wavenumber, i.e., the hyperdiffusion effect.
  This suggests that low-pass filtering strongly dampens short waves and preserves long waves; furthermore, with the increase in the number of filtering points, low-pass filtering better preserves microscale turbulence and mesoscale zonal structures in the non-aliased region. 
  
  Meanwhile, Eq.~\eqref{chik} can be conveniently used to estimate the numerical dissipation strength.
  For example, for typical parameters of a Tokamak device and setups of a simulation, the equivalent diffusion coefficients $\chi(k)$ corresponding to modes with different wavenumbers are presented in Fig. 2.
  Comparing the artificial dissipation introduced by low-pass filtering with the neoclassical thermal conductivity $\chi_{\text{neo}}$ induced by turbulent anomalous transport in the Tokamak plasma, it can be seen that the differential format with $N_{p} = 5$ destroys zonal structures significantly on mesoscales, but $\chi(k)$ for the $N_{p} = 9$ format approach the neoclassical level until close to the microscales region.
  Thus the five-point low-pass filtering used by part of the nonlinear gyrokinetic turbulence code~\cite{maeyama2013numerical} is difficult to ensure the reliability of the turbulence evolution during long-time simulations.
  \begin{figure}[htbp]
      \centering
      \includegraphics[width=0.9\textwidth]{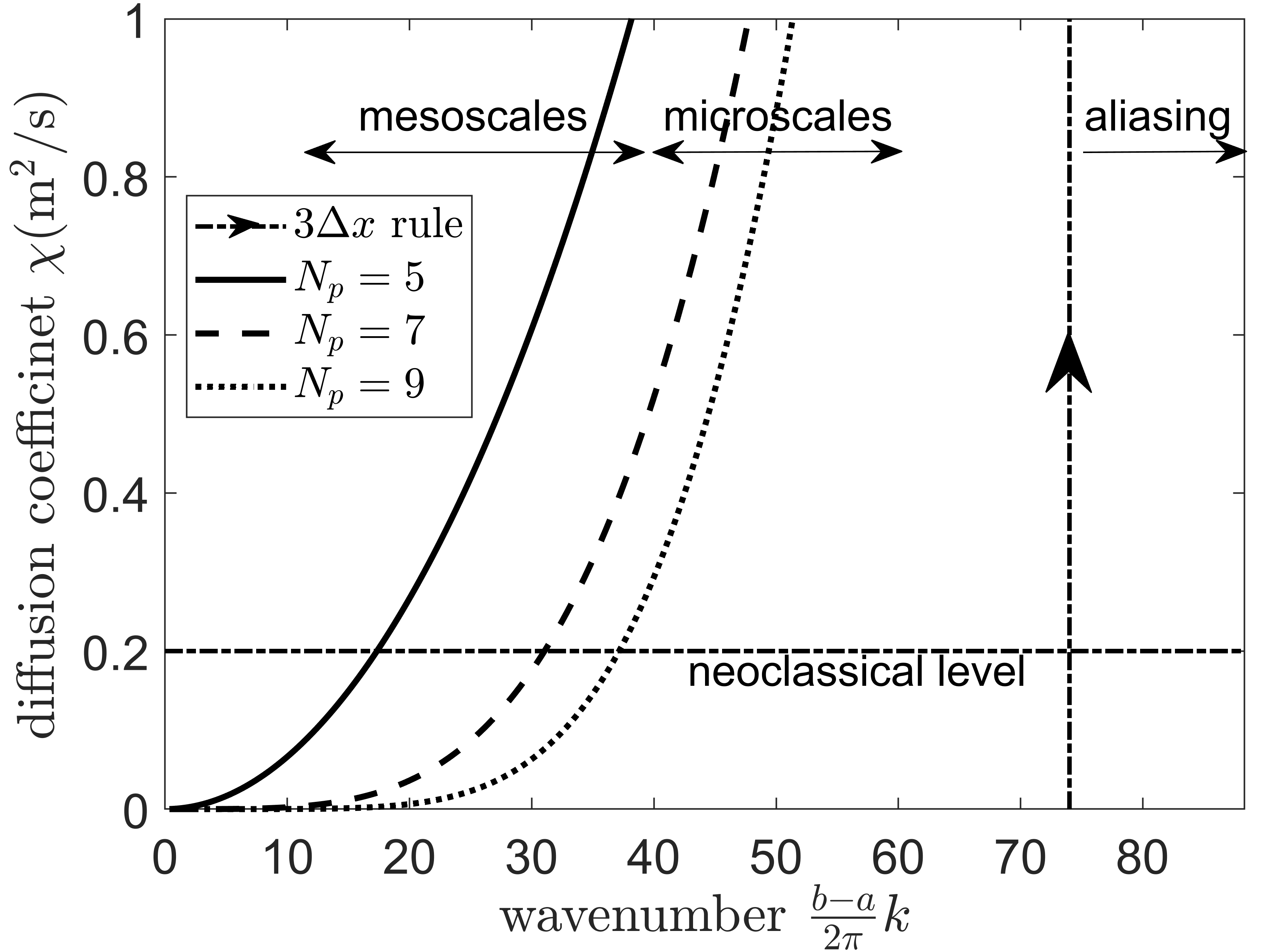}
      \caption{Equivalent diffusion coefficients $\chi(k)$ of the $3\Delta x$ rule (vertical dash-dotted) and the symmetric Pade-type discrete filters with five (solid), seven (dashed), nine (dotted) filter points corresponding respectively to cases 1,2,3 given in Table I. 
      The horizontal dash-dotted line denotes the neoclassical transport level $\chi_{\text{neo}} = 0.2 \mathrm{m^{2}/s}$. The mesoscale limit corresponds approximately to $6\Delta$. Here, $\Delta=0.6043 \mathrm{m}/222$, $N_{f,0} \approx 7.24\times 10^{6}$.}
      \label{fig:diffusion}
  \end{figure}

  \subsection{Boundary treatment}
  Implementation of the filtering schemes on domains with non-periodic boundaries requires the near boundary nodes to be treated separately.
  Different boundary conditions are applied for complex geometries and realistic physics-based considerations.
  It is well known that the boundary conditions for the filtered field are not necessarily the same as those for the non-filtered field.
  For example, in toroidal Tokamak the so-called field-aligned coordinates $(r,\alpha,\theta)$ are commonly used, where $r$, $\alpha$, $\theta$ are the non-periodic radial, periodic toroidal, and poloidal directions respectively.
  To ensure that the order of accuracy for the filter close to the boundary can be the same or higher than that in the interior of the domain, and to ensure the global dependence of the low-pass filtering at the boundary, a Pade-type compact formula is still chosen near the inner boundary ($r = 0$), which is written as
  \begin{equation}
      \phi_{1} + \underline{b}_{1}\phi_{1+1} = \sum_{n=-N_{ph}}^{+N_{ph}} \underline{c}_{n}\psi_{1+n}.
  \end{equation}
  The underline denotes the filter coefficients near the boundary. 
  Only the one-sided compact format is retained compared to Eq.~\eqref{Pade-type}, due to that the toroidal Fourier component of one field, $f_{n}(r,\theta)$, is a binary function about the non-periodic radial ($r$) and poloidal ($\theta$) directions, and the extension of the radial inner boundary involves the coupling to the poloidal direction. 
  Simultaneously, $\{\underline{c}_{n}\}$ lose symmetry.
  The Fourier transform $G(K)$ at the boundary is given by
  \begin{equation}
      G(K) = \sum_{n=-N_{ph}}^{+N_{ph}} \underline{c}_{n} \mathrm{e}^{\mathrm{i}nK}  \bigg/ \left(1+\underline{b}_{1}\mathrm{e}^{\mathrm{i}K}\right).
  \end{equation}
  
  To determine the filter coefficients, the constraints near the boundary are the same as those inside the domain, i.e., the highest order of accuracy is satisfied in the long-wave region, the response function decreases to zero at Nyquist wavenumber, and the last artificial constraint used to control the filter strength.
  The detailed derivation are described in appendix~\ref{derivation}.

\section{Nonlinear gyrokinetic simulations with the low-pass filters} \label{Simulations}
  In this section, applicability of the low-pass filter is examined in nonlinear turbulence simulations by using a continuum code, NLT~\cite{YeJCP16,XuPoP17,DaiCPC19}, to solve the time evolution of the nonlinear gyrokinetic equations in the Vlasov-Possion system.
  The NonLinear Turbulence (NLT) code used in this work, which adopts the field-aligned coordinates to solve the self-consistent evolution of electrostatic turbulence in Tokamake plasma, is based on the numerical Lie-transform method~\cite{WangPoP12,WangPoP13} derived from the I-transform theory.
  In the numerical Lie-transform method, the perturbed motion of the gyro-center is decoupled from the unperturbed motion; therefore, the evolution of distribution function along the unperturbed orbit is computed by the characteristic method and a four-dimension interpolation method~\cite{XiaoCCP17}, while the effects of the perturbed motion are evaluated by the numerical Lie-transform method which needs only to integrate the perturbed potential along the unperturbed orbit.
  The NLT code solves the time evolution of the gyrokinetic Vlasov equation by using finite difference methods for phase spatial discretization and the fourth-order Runge-Kutta scheme for time integration, and has been verified by various previous works, such as benchmarks of Rosenbulenth-Hinton test and the Cyclone base test of linear ITG mode~\cite{YeJCP16}, the nonlinear relaxation of ITG turbulence~\cite{DaiCPC19}, and the long-time nonlinear global turbulence evolution~\cite{wang2024self}.

  \subsection{Nonlinear relaxation of ITG turbulence}
  In this subsection, we focus on the nonlinear relaxation of ITG turbulence with kinetic deuterium and adiabatic electrons, and a general widely investigated "Cyclone DIII-D base case parameter set" without collision and source is used~\cite{dimits2000comparisons}.
  The main plasma parameters are set to be the mass ratio of electron to deuterium $m_{e}/m_{i}=1/3672$, the major radius in the  magnetic axis $R_{0} = 1.6714\mathrm{m}$, the minor radius of plasmas $a_{0} = 0.6043\mathrm{m}$, the magnetic field at the magnetic axis $B_{0} = 1.9\mathrm{T}$.
  The safety factor, temperature, and density profiles are set as
  \begin{subequations}
      \begin{align}
          q(r) &= 0.86 - 0.16\frac{r}{a_{0}} + 2.45 \left( \frac{r}{a_{0}} \right)^{2}, \\
          T(r) &= T_{0} \mathrm{exp} \left[ -\kappa_{T}\frac{a_{0}}{R_{0}}\Delta_{T} \mathrm{tanh}\left( \frac{(r-r_{0})/a_{0}}{\Delta_{T}} \right) \right], \\
          N(r) &= N_{0} \mathrm{exp} \left[ -\kappa_{N}\frac{a_{0}}{R_{0}}\Delta_{N} \mathrm{tanh}\left( \frac{(r-r_{0})/a_{0}}{\Delta_{N}} \right) \right],
      \end{align}
  \end{subequations}
  with $T_{0}=1.9693\mathrm{keV}$, $N_{0}=10^{19} \mathrm{m^{-3}}$, $\kappa_{T}=6.9589$, $\kappa_{N}=2.232$, $\Delta_{T}=\Delta_{N}=0.3$, $r_{0}=0.5a_{0}$.
  $r$ is the non-periodic radial direction used to examine the low-pass filtering.
  $\tau=T_{e,0}/T_{i,0}=1$ is assumed with $T_{i,0}$ and $T_{e,0}$ the initial ion and electron temperature, respectively.
  Hereafter, the normalizations of radius in unit of $a_{0}$ and time in unit of $R_{0}/c_{s}$ with the ion sound speed $c_{s}=\sqrt{T_{0}/m_{i}}$.
  In this work, the simulation domain is $r/a_{0}\in [0,1]$, $\theta\in [-\mathrm{\pi},\mathrm{\pi}]$, $\alpha \in [0,2\mathrm{\pi}]$, $v_{\parallel}/c_{s} \in [-6,6], \mu B_{0}/T_{0} \in [0, 6^{2}\sqrt{2}]$ with $v_{\parallel}$ and $\mu$ the parallel velocity and magnetic moment, respectively.
  In NLT code, $\mu$ is discretized according to the Gauss-Legendre formula, while the other variables and discretized uniformly.
  For nonlinear runs, we employ $222\times 190 \times 16 \times 96 \times 16$ grid points in $(r,\alpha,\theta,v_{\parallel},\mu)$.
  The time step size is set to be $\Delta t=2\tau_{c}$ with the ion gyro-period $\tau_{c}=2\mathrm{\pi} m_{i}/eB_{0}$ and the elementary charge $e$.

  \begin{figure}[htbp]
      \centering
      \includegraphics[width=0.9\textwidth]{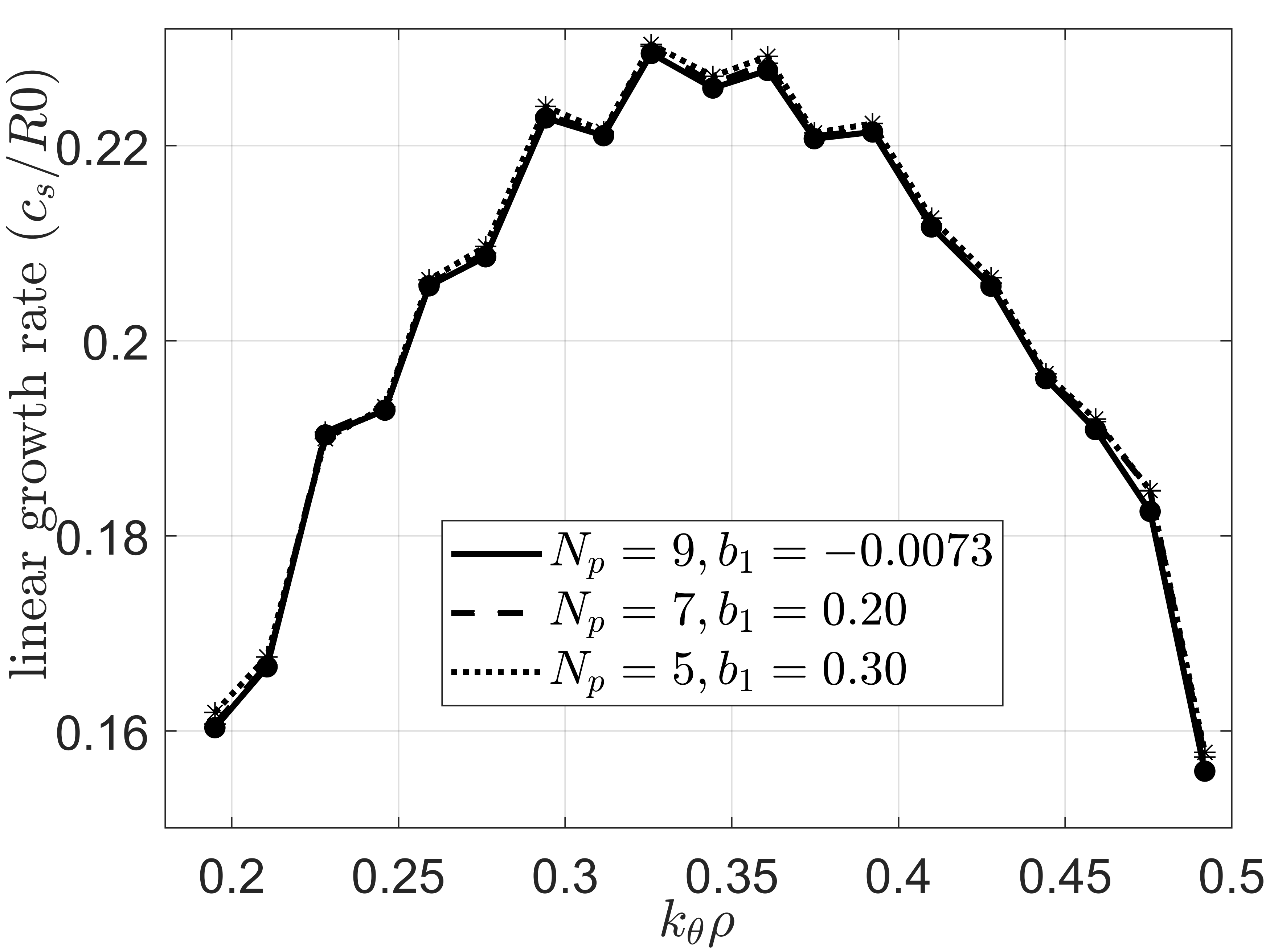}
      \caption{A set of linear growth rates versus the dimensionless wavenumber $k_{\theta}\rho$ with the filter points $N_{p}=5$ (dotted), $N_{p}=7$ (dashed), and $N_{p}=9$ (solid), where $k_{\theta}=\frac{m}{r}$ and $\rho=\frac{m_{i}c_{s}}{eB_{0}}$ are the perpendicular wavenumver and ion gyro-radius, respectively. Here, $m$ is the poloidal mode number.}
      \label{fig:gamma}
  \end{figure}

\begin{figure}[htbp]
      \centering
      \includegraphics[width=0.8\textwidth]{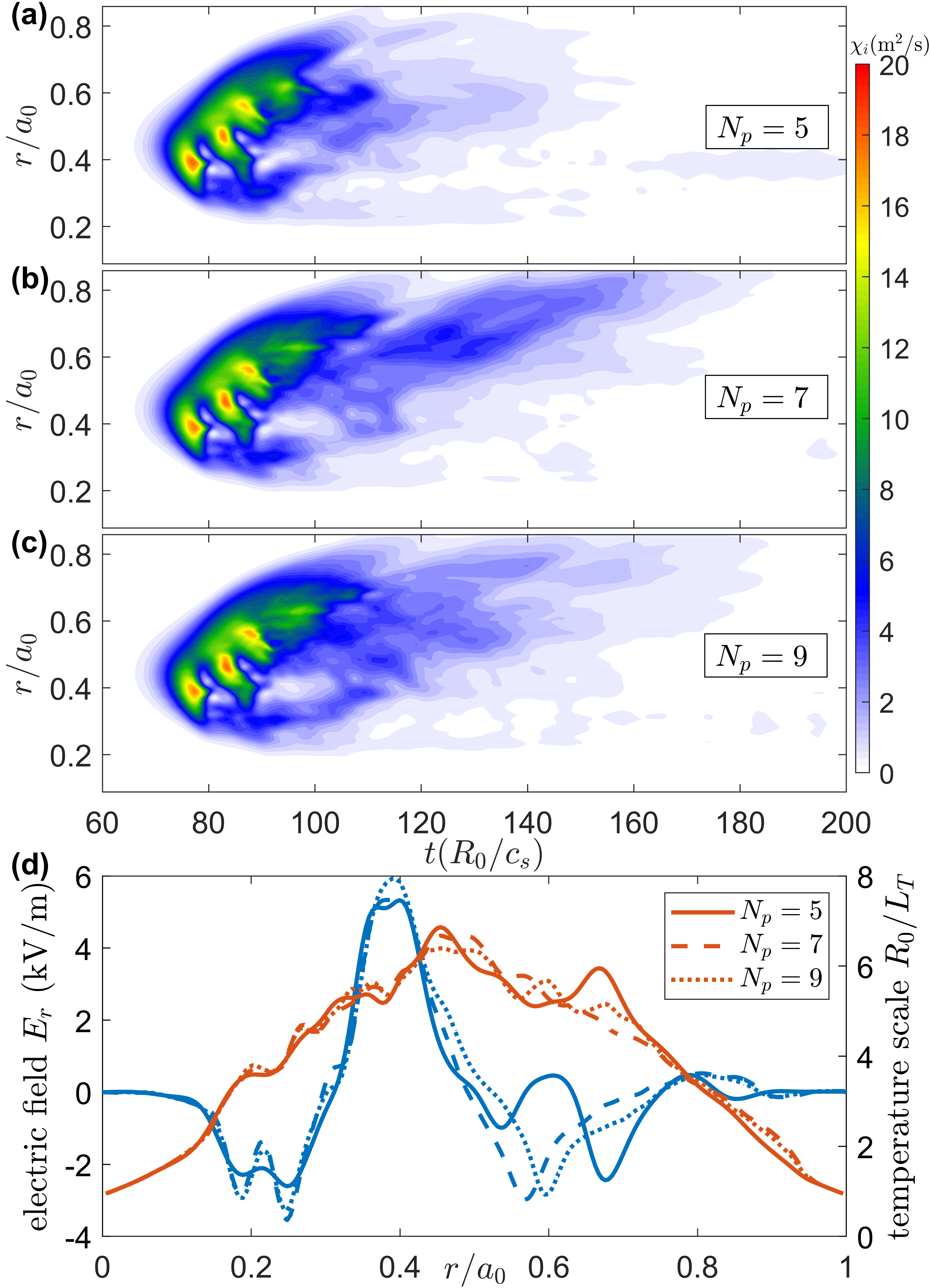}
      \caption{\textbf{a-c}: a set of two-dimensional evolution of ion thermal conductivity $\chi_{i}$ versus the simulation time and the radial position with the filter points (\textbf{a}) $N_{p}=5$, (\textbf{b}) $N_{p}=7$, and (\textbf{c}) $N_{p}=9$. The three low-pass filtered response functions correspond to the solid, dashed, and dotted lines in Fig.~\ref{fig:response}, respectively. \textbf{d}: zonal radial electric field $E_{r}$ (blue) and ion temperature scale length $R_{0}/L_{T}$ (orange) versus radial direction at $t=190 R_{0}/c_{s}$.}
      \label{fig:format}
  \end{figure}

  The filtering module of the NLT code utilizes the $3\Delta x$ rule~\cite{orszag1971elimination} for the periodic toroidal and polar directions in the Tokamak plasma to avoid aliasing occurrences, and low-pass filtering is applied in the non-periodic radial direction.
  As can be seen in Fig.~\ref{fig:gamma}, the filter format with different points has almost no effect on the linear growth rate of the ITG modes, yet in Fig.~\ref{fig:format}(a-c) the nonlinear evolution of the turbulence is different with different filter formats.
  Overall, the ITG turbulence gets stronger during the relaxation when the number of filter points is larger.
  
  By comparing Fig.~\ref{fig:gamma} and Fig.~\ref{fig:format}, it is observed that the dissipative effect of low-pass filtering on the ITG turbulence evolution is not directly on the turbulence itself, but rather destroys the zonal structures on mesoscales, such as temperature profiles or zonal flows.
  In Fig.~\ref{fig:format}(d), the zonal radial electric field obtained from the simulation with low-pass filtering for $N_{p}=5$ is significantly different from that obtained for $N_{p}=7$ and $N_{p}=9$.
  Specifically, the disappearance of the short-wave undulations near $r=0.2a_{0}$ indicates that the five-point format filter causes considerable damage to zonal structures, while the non-convergence in the region of $0.6a_{0}-0.7a_{0}$ suggests that the five-point low-pass filtering is not acceptable in nonlinear simulations of long-duration ITG turbulence.

  \subsection{Dissipation of zonal structures by low-pass filters}
  In this subsection, we illustrate that the effect of low-pass filtering on the nonlinear evolution of turbulence is achieved indirectly brought by dissipating the structure of the neighboring nonlinear equilibrium by designing simulations with continuous dissipating the zonal structures through the low-pass filtering.
  Long-lived zonal structures, which are nonlinearly excited by transient fluctuating fields and become part of the equilibrium state in Tomakak plasmas, connect the macroscopic and microscopic scales, and in particular exemplify the slow evolution in the macroscale plasma profile on the energy transport time.
  In this sense, the radial low-pass filtering needs to be treated carefully.

  To investigate the dissipation of stable zonal structures and the resulting distortion of the turbulence by radial low-pass filtering, we first turn off the evolution of fluctuations and zonal structures after the plasma turbulence quenching, but keep the low-pass filtering on zonal structures, which is achieved by applying the radial low-pass filtering to the toroidally averaged distribution function for up to $2.2\mathrm{ms}$.
  This time scale is intermediate between the turbulent single burst and the energy transport time, and is also the interval between the intermittent bursts of turbulence or the characteristic time of zonal structures when the core plasma confinement is improved~\cite{wang2024self}.
  We then turn on the nonlinear evolution of the plasma again to observe the subsequent morphology of the turbulence and zonal structures in two different filter formats.

  \begin{figure}[htbp]
      \centering
      \includegraphics[width=0.8\textwidth]{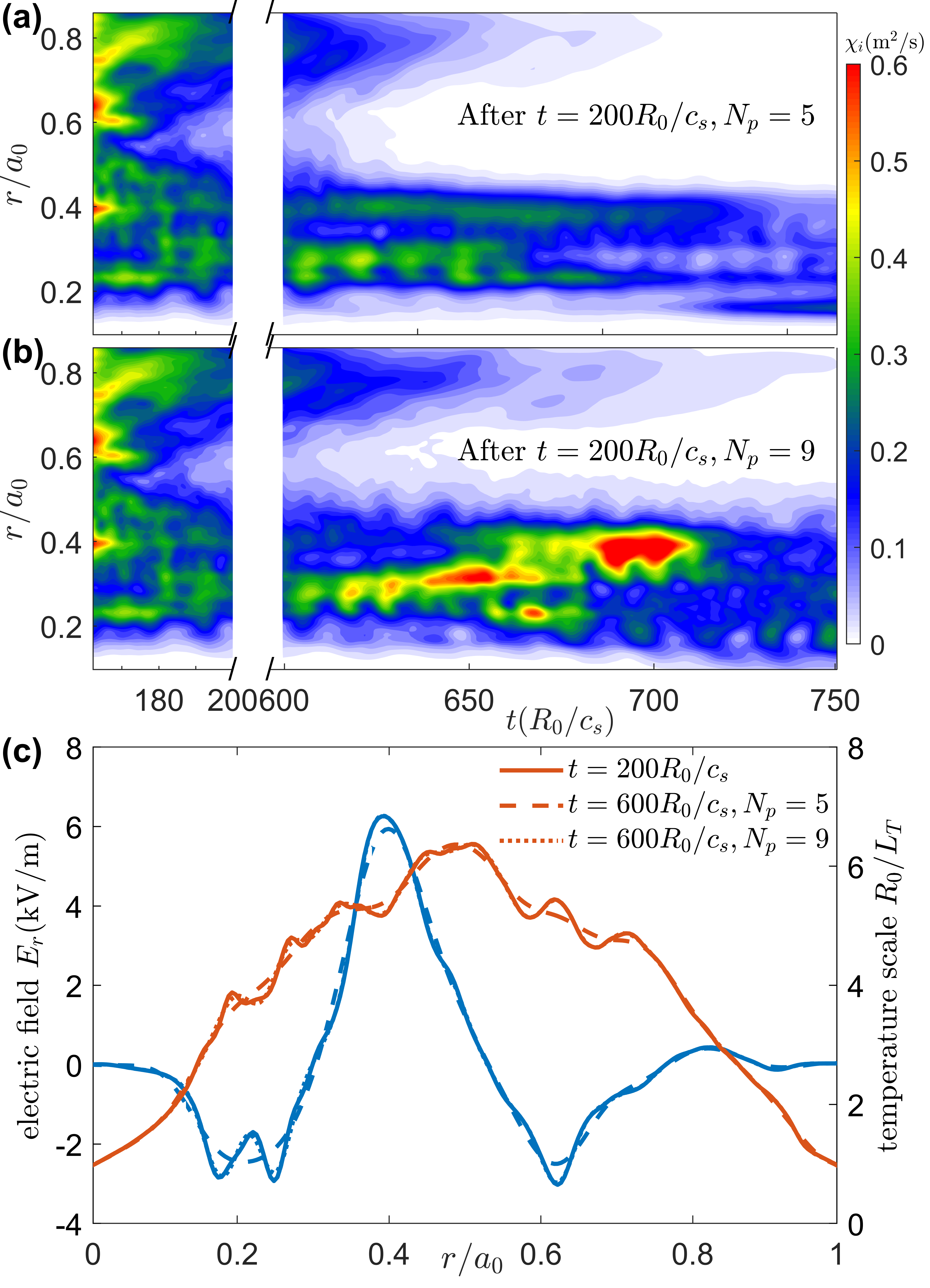}
      \caption{\textbf{a-b}: a set of two-dimensional evolution of ion thermal conductivity $\chi_{i}$ versus the simulation time and the radial position, with the filter points $N_{p}=9$ for both before $t=200 R_{0}/c_{s}$, while (\textbf{a}) $N_{p}=5$ and (\textbf{b}) $N_{p}=9$ after $t=200 R_{0}/c_{s}$. \textbf{c}: zonal radial electric field $E_{r}$ (blue) and ion temperature scale length $R_{0}/L_{T}$ (orange) versus radial direction.}
      \label{fig:chi}
  \end{figure}

  From Fig.~\ref{fig:chi}(a-b), we find that the turbulence evolution is exactly the same for both simulations for $N_{p} = 9$ before continuous filtering on zonal structures, but when two different formats of low-pass filtering, $N_{p} = 5$ and $N_{p} = 9$, are applied during continuous filtering and subsequent switching on of the nonlinear evolution, respectively, the ion thermal conductivity of the latter are significantly higher compared to the former.
  The reason for this discrepancy is determined to be the dissipation of the radial zonal structures by multiple low-pass filtering.
  Fig.~\ref{fig:chi}(c) shows that the application of low-pass filtering with nine-point format in long-time numerical simulations hardly affects zonal structures, e.g., the zonal radial electric field and temperature profiles, whereas the five-point format significantly disrupts the smaller-scale radial structure of zonal flows and smoothes out the undulation of the temperature scales.
  The five-point format modifies the neighboring nonlinear equilibrium~\cite{chen2007nonlinear,wang2024self}, which is manifested here by flattening the temperature profile, leading to a more stable turbulence evolution.

  Note that the characteristic dissipation time of the zonal structures, $t_{D}$, can be quickly estimated by using Eq.~\eqref{final_constraint} and an approximate expression for the response function, Eq.~\eqref{response_approximate}.
  Taking zonal flows in the region near $r=0.2a_{0}$ in Fig.~\ref{fig:chi} as an example, $t_{D}$ (normalized by $R_{0}/c_{s}$) for a radial undulation with its radial wavelength $\lambda_{r} \approx 0.05a_{0}$ overdamped to $\beta = 1/\mathrm{e}$ of the initial amplitude is
  \begin{equation} \label{t_D}
      t_{D} = \frac{1}{\epsilon\cdot K^{N_{p}-1}} \cdot \frac{c_{s}}{R_{0}}\Delta t \approx 
      \begin{cases}
          16 \qquad ~\text{for}~ N_{p}=5; \\
          602 \qquad \text{for}~ N_{p}=9.
      \end{cases}
  \end{equation}
  Here, $K=\frac{2\mathrm{\pi} a_{0}}{N_{x}\lambda_{r}} \approx 0.566$; $\frac{c_{s}}{R_{0}}\Delta t \approx 0.0254$; $\epsilon = \epsilon(N_{p},b_{1}) \approx 0.0154 / 0.0040$, shown in Eq.~\eqref{epsilon} and Table I, respectively correspond to $N_{p}=5$ and $N_{p}=9$. 
  From Eq.~\eqref{t_D}, it can be seen that the characteristic diffusion time of zonal flows in the case of $N_{p} = 9$ is much longer than the characteristic time of its own stable existence, so that the low-pass filtering does not destroy zonal structures in long-time gyrokinetic simulations. 
  In contrast, when $N_{p}=5$, the low-pass filtering significantly dissipates zonal structures in a short time (even less than the turbulence outburst time), which is obviously unacceptable.

\section{Conclusion} \label{Conclution}

  In this work we have discussed the application of low-pass filtering with finite difference format to gyrokinetic simulations.
  Overall, the desirable bandpass characteristics of low-pass filters are close enough to Fourier filters to eliminate aliasing errors without affecting the numerical results of the longwave to be solved.
  Thus, this work first lists the generalised difference scheme that ensures the highest order of numerical accuracy in the longwave region, which can be used for long-time gyrokinetic simulations with variable time steps, and derives the equivalent diffusion coefficients introduced by the low-pass filter in the longwave region.
  It is shown analytically that this dissipation coefficient is proportional to $K^{N_{p}-3}$, so that for a given normalized wavelength, the higher the number of filtering points, the weaker the dissipation of micro-scale turbulence and meso- to macro-scale zonal structures in the non-aliased region.
  This is consistent with the previous opinion that low-pass filtering becomes a better approximation to shape cutoff filtering with the increase in the number of filter points~\cite{vasilyev1998general}.
  The novelty of this derivation is that the dissipation strength of low-pass filtering is given quantitatively, and to some extent the estimation of the equivalent diffusion coefficients can be viewed as a convergence test for low-pass filtering in nonlinear turbulence simulations; in particular, this type of pre-estimation is of value in massively parallel gyrokinetic simulations consuming large amounts of computational resources.

  We then apply the optimised low-pass filtering to the NLT code.
  The results show that the adoption of different formats of low-pass filtering has almost no effect on the linear growth rate of the ITG modes in Tokamak plasmas, but the nonlinear evolution of the ITG turbulence diverges considerably, and zonal structures driven nonlinearly by turbulence, such as zonal flows and temperature profiles, are similarly differently distributed in the radial direction.
  These inconsistent results suggest that the five-point format low-pass filtering is inappropriate while the nine-point format is appropriate in nonlinear gyrokinetic simulations, which is consistent with the previous findings given from the estimation of the equivalent diffusion coefficients.

  Finally, a set of comparisons of successive low-pass filtering of zonal structures demonstrates that the effect of low-pass filtering on system reliability is brought by destroying the mesoscale zonal structures.
  These long-lived zonal structures, as an important piece in the neighboring nonlinear equilibrium, are continuously changed by radial low-pass filtering.
  In the picture shown in this work, the flattening of the mesoscopic temperature scale length by inappropriate low-pass filtering subsequently changes the evolution of the ITG turbulent.
  These pictures further exemplify the importance of mesoscale zonal structures in nonlinear turbulent transport in Tokamak plasmas, and also require that low-pass filtering should be handled carefully in numerical simulation.

  This work was supported by the National MCF Energy R\&D Program of China under Grant No. 2019YFE03060000, and the National Natural Science Foundation of China under Grant No. 12075240.

\appendix
\section{Detailed derivation of finite difference schemes} \label{derivation}

  In this appendix, we derive in detail how to determine the discrete low-pass filter coefficients.
  In the case of symmetric Pade-type filters, the constraint of zero-order Taylor series of Eq.~\eqref{response} is requires that 
  \begin{equation} \label{zero}
      c_{0} + \sum_{n=1}^{N_{ph}} 2c_{n} = 1 + 2b_{1}.
  \end{equation}
  The Taylor series of higher-order up to $2(N_{ph}-1)$ order further require that
  \begin{equation} \label{higher}
      \sum_{n=1}^{N_{ph}} c_{n}\cdot n^{2l} = b_{1}, \quad l = 1, ..., N_{ph}-1.
  \end{equation}
  The desirable constraint that $G(\mathrm{\pi}) = 0$ at the Nyquist wavenumber requires that
  \begin{equation} \label{Nyquist}
      c_{0} + \sum_{n=1}^{N_{ph}} (-1)^{n} 2c_{n} = 0.
  \end{equation}
  Combining Eqs.~(\ref{zero}-\ref{Nyquist}), one can represents the difference coefficients $\{c_{n}\}$ as a one-dimension family of the compact coefficient ($b_{1}$), containing the zero $\{c_{n,0}\}$ and primary $\{c_{n,1}\}$ terms.
  Examples of relations for symmetric Pade-type filters with $N_{p}=5,7,9$ are given in Table I.
  
  \begin{table*}[bp] \label{table:coefficient}
  \renewcommand\arraystretch{1.5}
    \centering
    \begin{tabular}{cccccccc}
    \hline
        Case & $N_{p}$ & $c_{0}$ & $c_{1}$ & $c_{2}$ & $c_{3}$ & $c_{4}$ & $\epsilon$ \\ \hline
        1 & 5 & $\frac{5+6b_{1}}{8}$ & $\frac{1+2b_{1}}{4}$ & $\frac{-1+2b_{1}}{16}$ & ~ & ~ & $\frac{1-2b_{1}}{16(1+2b_{1})}$ \\ %\hline
        2 & 7 & $\frac{11+10b_{1}}{16}$ & $\frac{15+34b_{1}}{64}$ & $\frac{-3+6b_{1}}{32}$ & $\frac{1-2b_{1}}{64}$ & ~ & $\frac{45-82b_{1}}{2880(1+2b_{1})}$ \\ %\hline
        3 & 9 & $\frac{93+70b_{1}}{128}$ & $\frac{7+18b_{1}}{32}$ & $\frac{-7+14b_{1}}{64}$ & $\frac{1-2b_{1}}{32}$ & $\frac{-1+2b_{1}}{256}$ & $\frac{315-634b_{1}}{80640(1+2b_{1})}$ \\ \hline
    \end{tabular}
    \caption{Values of difference coefficients and truncation errors for symmetric Pade-type filters with different point numbers}
  \end{table*}

  The first unmatched coefficient (here is the Taylor series of ($N_{p}-1$) order) determines the formal truncation error of low-pass filters.
  Under the constraint that the highest order accuracy is satisfied in the long wave region, there is a relationship between the amplitude before and after low-pass filtering,
  \begin{equation}
    \phi_{K} = \left[1 - \epsilon \cdot K^{N_{p}-1} + \mathcal{o}(K^{N_{p}-1}) \right]\psi_{K}.
  \end{equation}
  To determine the truncation error coefficient $\epsilon$, perform the Taylor series expansion of Eq.~\eqref{response},
  \begin{equation}
  \begin{split}
      \left[ 1+2b_{1} + \sum_{l=1}^{\infty} \frac{2\cdot (-1)^{l}}{(2l)!} b_{1}\cdot K^{2l} \right] (1-\epsilon\cdot K^{N_{p}-1})\psi_{K} \\
      = \left[ c_{0}+\sum_{n=1}^{N_{ph}}2c_{n} + \sum_{l=1}^{\infty}\frac{2\cdot (-1)^{l}}{(2l)!} \sum_{l=1}^{N_{ph}} c_{n}\cdot (nK)^{2l} \right]\psi_{K}.
  \end{split}
  \end{equation}
  According to the $({N_{p}-1)}$ order coefficient, one obtains
  \begin{equation} \label{epsilon}
      \epsilon = \frac{-2\cdot (-1)^{N_{ph}}}{(1+2b_{1})\cdot (2N_{ph})!} \left( \sum_{n=1}^{N_{ph}}c_{n}\cdot n^{2N_{ph}}-b_{1} \right).
  \end{equation}
  Examples of truncation errors for symmetric Pade-type filters with $N_{p}=5,7,9$ are also given in Table I.

\begin{figure}[htp]
      \centering
      \includegraphics[width=0.9\textwidth]{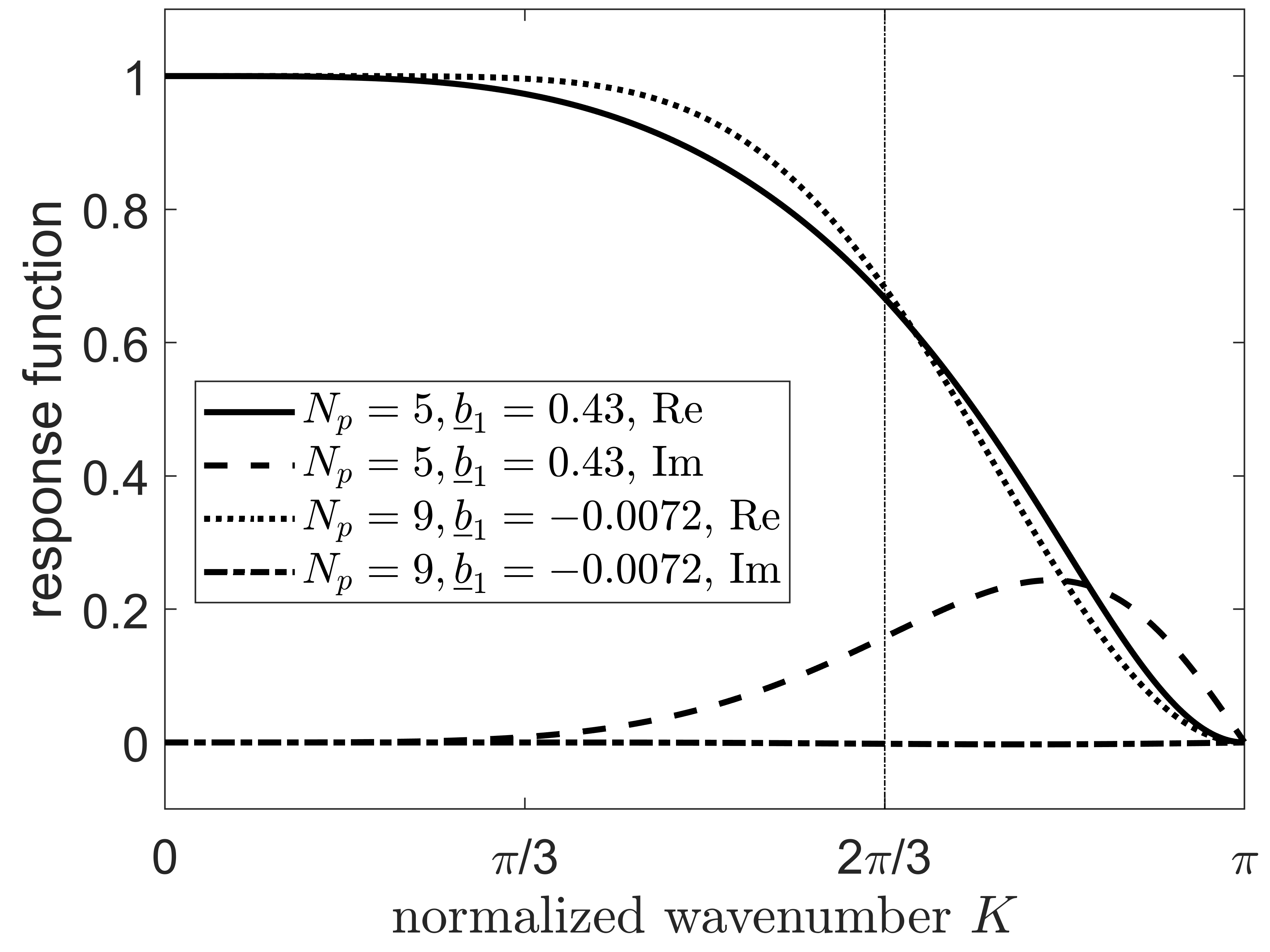}
      \caption{Fourier transform $G(K)$ of the Pade-type discrete filters near the boundary with five (solid for real part, dashed for imaginary part) and nine (dotted for real part, horizontal dash-dotted for imaginary part) filter points corresponding respectively to cases 4 and 6 given in Table II, which share the same value, $|G(K_{0})|$, at $K_{0} = 2\mathrm{\pi}/3$; the parameter in Eq.~\ref{last_constraint} is $N_{f}=6, \beta=0.1$. The right side of the vertical dash-dotted line denotes the aliasing area.}
      \label{fig:response_boundary}
  \end{figure}

  In the case of Pade-type filters near the boundary, conditions~(\ref{zero}-\ref{Nyquist}) can be rewritten as
  \begin{subequations} \label{boundary}
      \begin{align}
            \sum_{n=-N_{ph}}^{+N_{ph}} &\underline{c}_{n} = 1 + \underline{b}_{1}; \\
            \sum_{n=-N_{ph}}^{+N_{ph}} &\underline{c}_{n}\cdot n^{l} = \underline{b}_{1}, \quad l = 1, ..., N_{p}-2; \\
            \sum_{n=-N_{ph}}^{+N_{ph}} &(-1)^{n} \underline{c}_{n} = 0.
      \end{align}
  \end{subequations}
  These conditions give the maximum filter support for a Pade-type low-pass filter in order to provide the highest accuracy of low-pass filtering in the long-wave region.
  Examples of weights for conditions~\eqref{boundary} are given in Table II, and associated Fourier response functions for some of these filters with $N_{p}=5,9$ are presented in Fig~\ref{fig:response_boundary}.
  \begin{table*}[!ht] \label{table:boundary}
  \renewcommand\arraystretch{1.5}
    \centering
    \begin{tabular}{ccccccccccccc}
    \hline
        Case & $N_{p}$ & $\underline{c}_{-4}$ & $\underline{c}_{-3}$ & $\underline{c}_{-2}$ & $\underline{c}_{-1}$ & $\underline{c}_{0}$ & $\underline{c}_{1}$ & $\underline{c}_{2}$ & $\underline{c}_{3}$ & $\underline{c}_{4}$ \\ \hline
        4 & 5 & ~ & ~ & $\frac{-1+\underline{b}_{1}}{16}$ & $\frac{1-\underline{b}_{1}}{4}$ & $\frac{5+3\underline{b}_{1}}{8}$ & $\frac{1+3\underline{b}_{1}}{4}$ & $\frac{-1+\underline{b}_{1}}{16}$ & ~ & ~ \\ %\hline
        5 & 7 & ~ & $\frac{1-\underline{b}_{1}}{64}$ & $\frac{-3+3\underline{b}_{1}}{32}$ & $\frac{15-15\underline{b}_{1}}{64}$ & $\frac{11+5\underline{b}_{1}}{16}$ & $\frac{15+49\underline{b}_{1}}{64}$ & $\frac{-3+3\underline{b}_{1}}{32}$ & $\frac{1-\underline{b}_{1}}{64}$ & ~ \\ %\hline
        6 & 9 & $\frac{-1+\underline{b}_{1}}{256}$ & $\frac{1-\underline{b}_{1}}{32}$ & $\frac{-7+7\underline{b}_{1}}{64}$ & $\frac{7-7\underline{b}_{1}}{32}$ & $\frac{93+35\underline{b}_{1}}{128}$ & $\frac{7+25\underline{b}_{1}}{32}$ & $\frac{-7+7\underline{b}_{1}}{64}$ & $\frac{1-\underline{b}_{1}}{32}$ & $\frac{-1+\underline{b}_{1}}{256}$ \\ \hline
    \end{tabular}
    \caption{Values of difference coefficients for Pade-type filters near the boundary with different point numbers}
  \end{table*}

  The last constraint of Pade-type low-pass filters near the boundary is
  \begin{equation} \label{last_constraint}
      |G(K_{0})|^{N_{f}} = \beta.
  \end{equation}
  The truncation error of Pade-type low-pass filters near the boundary is
  \begin{equation} \label{truncation_boundary}
      \underline{\epsilon} = \frac{-\mathrm{i}^{N_{p}-1}}{(1+\underline{b}_{1})(N_{p}-1)!} \left[ \sum_{n=-N_{ph}}^{+N_{ph}} \underline{c}_{n}\cdot n^{N_{p}-1}-\underline{b}_{1} \right].
  \end{equation}
  From Eq.~\eqref{truncation_boundary}, it can be seen that the truncation error is real when the number of filtering points is odd, such as $N_{p}=5,7,9$, so that no additional dispersion is introduced by low-pass filtering in the long wavelength region, which can be observed from Fig.~\ref{fig:response_boundary} where the imaginary part of the response function in  is nearly zero for $K < \mathrm{\pi}/3$.

% If in two-column mode, this environment will change to single-column format so that long equations can be displayed. 
% Use only when necessary.
%\begin{widetext}
%$$\mbox{put long equation here}$$
%\end{widetext}

% Figures should be put into the text as floats. 
% Use the graphics or graphicx packages (distributed with LaTeX2e).
% See the LaTeX Graphics Companion by Michel Goosens, Sebastian Rahtz, and Frank Mittelbach for examples. 
%
% Here is an example of the general form of a figure:
% Fill in the caption in the braces of the \caption{} command. 
% Put the label that you will use with \ref{} command in the braces of the \label{} command.
%
% \begin{figure}
% \includegraphics{}%
% \caption{\label{}}%
% \end{figure}

% Tables may be be put in the text as floats.
% Here is an example of the general form of a table:
% Fill in the caption in the braces of the \caption{} command. Put the label
% that you will use with \ref{} command in the braces of the \label{} command.
% Insert the column specifiers (l, r, c, d, etc.) in the empty braces of the
% \begin{tabular}{} command.
%
% \begin{table}
% \caption{\label{} }
% \begin{tabular}{}
% \end{tabular}
% \end{table}

% If you have acknowledgments, this puts in the proper section head.
%\begin{acknowledgments}
% Put your acknowledgments here.
%\end{acknowledgments}

% Create the reference section using BibTeX:
\bibliography{ref}

%merlin.mbs aipnum4-1.bst 2010-07-25 4.21a (PWD, AO, DPC) hacked
%Control: key (0)
%Control: author (8) initials jnrlst
%Control: editor formatted (1) identically to author
%Control: production of article title (0) allowed
%Control: page (1) range
%Control: year (1) truncated
%Control: production of eprint (0) enabled
\begin{thebibliography}{39}%
\makeatletter
\providecommand \@ifxundefined [1]{%
 \@ifx{#1\undefined}
}%
\providecommand \@ifnum [1]{%
 \ifnum #1\expandafter \@firstoftwo
 \else \expandafter \@secondoftwo
 \fi
}%
\providecommand \@ifx [1]{%
 \ifx #1\expandafter \@firstoftwo
 \else \expandafter \@secondoftwo
 \fi
}%
\providecommand \natexlab [1]{#1}%
\providecommand \enquote  [1]{``#1''}%
\providecommand \bibnamefont  [1]{#1}%
\providecommand \bibfnamefont [1]{#1}%
\providecommand \citenamefont [1]{#1}%
\providecommand \href@noop [0]{\@secondoftwo}%
\providecommand \href [0]{\begingroup \@sanitize@url \@href}%
\providecommand \@href[1]{\@@startlink{#1}\@@href}%
\providecommand \@@href[1]{\endgroup#1\@@endlink}%
\providecommand \@sanitize@url [0]{\catcode `\\12\catcode `\$12\catcode `\&12\catcode `\#12\catcode `\^12\catcode `\_12\catcode `\%12\relax}%
\providecommand \@@startlink[1]{}%
\providecommand \@@endlink[0]{}%
\providecommand \url  [0]{\begingroup\@sanitize@url \@url }%
\providecommand \@url [1]{\endgroup\@href {#1}{\urlprefix }}%
\providecommand \urlprefix  [0]{URL }%
\providecommand \Eprint [0]{\href }%
\providecommand \doibase [0]{http://dx.doi.org/}%
\providecommand \selectlanguage [0]{\@gobble}%
\providecommand \bibinfo  [0]{\@secondoftwo}%
\providecommand \bibfield  [0]{\@secondoftwo}%
\providecommand \translation [1]{[#1]}%
\providecommand \BibitemOpen [0]{}%
\providecommand \bibitemStop [0]{}%
\providecommand \bibitemNoStop [0]{.\EOS\space}%
\providecommand \EOS [0]{\spacefactor3000\relax}%
\providecommand \BibitemShut  [1]{\csname bibitem#1\endcsname}%
\let\auto@bib@innerbib\@empty
%</preamble>
\bibitem [{\citenamefont {Tajima}(2018)}]{tajima2018computational}%
  \BibitemOpen
  \bibfield  {author} {\bibinfo {author} {\bibfnamefont {T.}~\bibnamefont {Tajima}},\ }\href@noop {} {\emph {\bibinfo {title} {Computational plasma physics: with applications to fusion and astrophysics}}}\ (\bibinfo  {publisher} {CRC press},\ \bibinfo {year} {2018})\BibitemShut {NoStop}%
\bibitem [{\citenamefont {Deardorff}(1970)}]{deardorff1970numerical}%
  \BibitemOpen
  \bibfield  {author} {\bibinfo {author} {\bibfnamefont {J.~W.}\ \bibnamefont {Deardorff}},\ }\bibfield  {title} {\enquote {\bibinfo {title} {A numerical study of three-dimensional turbulent channel flow at large reynolds numbers},}\ }\href@noop {} {\bibfield  {journal} {\bibinfo  {journal} {J. Fluid Mech.}\ }\textbf {\bibinfo {volume} {41}},\ \bibinfo {pages} {453--480} (\bibinfo {year} {1970})}\BibitemShut {NoStop}%
\bibitem [{\citenamefont {Schumann}(1975)}]{schumann1975subgrid}%
  \BibitemOpen
  \bibfield  {author} {\bibinfo {author} {\bibfnamefont {U.}~\bibnamefont {Schumann}},\ }\bibfield  {title} {\enquote {\bibinfo {title} {Subgrid scale model for finite difference simulations of turbulent flows in plane channels and annuli},}\ }\href@noop {} {\bibfield  {journal} {\bibinfo  {journal} {J. Comput. Phys.}\ }\textbf {\bibinfo {volume} {18}},\ \bibinfo {pages} {376--404} (\bibinfo {year} {1975})}\BibitemShut {NoStop}%
\bibitem [{\citenamefont {Birdsall}\ and\ \citenamefont {Langdon}(2018)}]{birdsall2018plasma}%
  \BibitemOpen
  \bibfield  {author} {\bibinfo {author} {\bibfnamefont {C.~K.}\ \bibnamefont {Birdsall}}\ and\ \bibinfo {author} {\bibfnamefont {A.~B.}\ \bibnamefont {Langdon}},\ }\href@noop {} {\emph {\bibinfo {title} {Plasma physics via computer simulation}}}\ (\bibinfo  {publisher} {CRC press},\ \bibinfo {year} {2018})\BibitemShut {NoStop}%
\bibitem [{\citenamefont {Langdon}(1970)}]{langdon1970effects}%
  \BibitemOpen
  \bibfield  {author} {\bibinfo {author} {\bibfnamefont {A.~B.}\ \bibnamefont {Langdon}},\ }\bibfield  {title} {\enquote {\bibinfo {title} {Effects of the spatial grid in simulation plasmas},}\ }\href@noop {} {\bibfield  {journal} {\bibinfo  {journal} {J. Comput. Phys.}\ }\textbf {\bibinfo {volume} {6}},\ \bibinfo {pages} {247--267} (\bibinfo {year} {1970})}\BibitemShut {NoStop}%
\bibitem [{\citenamefont {Godfrey}(1974)}]{godfrey1974numerical}%
  \BibitemOpen
  \bibfield  {author} {\bibinfo {author} {\bibfnamefont {B.~B.}\ \bibnamefont {Godfrey}},\ }\bibfield  {title} {\enquote {\bibinfo {title} {Numerical cherenkov instabilities in electromagnetic particle codes},}\ }\href@noop {} {\bibfield  {journal} {\bibinfo  {journal} {J. Comput. Phys.}\ }\textbf {\bibinfo {volume} {15}},\ \bibinfo {pages} {504--521} (\bibinfo {year} {1974})}\BibitemShut {NoStop}%
\bibitem [{\citenamefont {Xu}\ \emph {et~al.}(2013)\citenamefont {Xu}, \citenamefont {Yu}, \citenamefont {Martins}, \citenamefont {Tsung}, \citenamefont {Decyk}, \citenamefont {Vieira}, \citenamefont {Fonseca}, \citenamefont {Lu}, \citenamefont {Silva},\ and\ \citenamefont {Mori}}]{xu2013numerical}%
  \BibitemOpen
  \bibfield  {author} {\bibinfo {author} {\bibfnamefont {X.}~\bibnamefont {Xu}}, \bibinfo {author} {\bibfnamefont {P.}~\bibnamefont {Yu}}, \bibinfo {author} {\bibfnamefont {S.~F.}\ \bibnamefont {Martins}}, \bibinfo {author} {\bibfnamefont {F.~S.}\ \bibnamefont {Tsung}}, \bibinfo {author} {\bibfnamefont {V.~K.}\ \bibnamefont {Decyk}}, \bibinfo {author} {\bibfnamefont {J.}~\bibnamefont {Vieira}}, \bibinfo {author} {\bibfnamefont {R.~A.}\ \bibnamefont {Fonseca}}, \bibinfo {author} {\bibfnamefont {W.}~\bibnamefont {Lu}}, \bibinfo {author} {\bibfnamefont {L.~O.}\ \bibnamefont {Silva}}, \ and\ \bibinfo {author} {\bibfnamefont {W.~B.}\ \bibnamefont {Mori}},\ }\bibfield  {title} {\enquote {\bibinfo {title} {Numerical instability due to relativistic plasma drift in em-pic simulations},}\ }\href@noop {} {\bibfield  {journal} {\bibinfo  {journal} {Comput. Phys. Commun.}\ }\textbf {\bibinfo {volume} {184}},\ \bibinfo {pages} {2503--2514} (\bibinfo {year} {2013})}\BibitemShut {NoStop}%
\bibitem [{\citenamefont {Huang}\ \emph {et~al.}(2016)\citenamefont {Huang}, \citenamefont {Zeng}, \citenamefont {Wang}, \citenamefont {Meyers}, \citenamefont {Yi},\ and\ \citenamefont {Albright}}]{huang2016finite}%
  \BibitemOpen
  \bibfield  {author} {\bibinfo {author} {\bibfnamefont {C.-K.}\ \bibnamefont {Huang}}, \bibinfo {author} {\bibfnamefont {Y.}~\bibnamefont {Zeng}}, \bibinfo {author} {\bibfnamefont {Y.}~\bibnamefont {Wang}}, \bibinfo {author} {\bibfnamefont {M.~D.}\ \bibnamefont {Meyers}}, \bibinfo {author} {\bibfnamefont {S.}~\bibnamefont {Yi}}, \ and\ \bibinfo {author} {\bibfnamefont {B.~J.}\ \bibnamefont {Albright}},\ }\bibfield  {title} {\enquote {\bibinfo {title} {Finite grid instability and spectral fidelity of the electrostatic particle-in-cell algorithm},}\ }\href@noop {} {\bibfield  {journal} {\bibinfo  {journal} {Comput. Phys. Commun.}\ }\textbf {\bibinfo {volume} {207}},\ \bibinfo {pages} {123--135} (\bibinfo {year} {2016})}\BibitemShut {NoStop}%
\bibitem [{\citenamefont {Maeyama}\ \emph {et~al.}(2013)\citenamefont {Maeyama}, \citenamefont {Ishizawa}, \citenamefont {Watanabe}, \citenamefont {Nakajima}, \citenamefont {Tsuji-Iio},\ and\ \citenamefont {Tsutsui}}]{maeyama2013numerical}%
  \BibitemOpen
  \bibfield  {author} {\bibinfo {author} {\bibfnamefont {S.}~\bibnamefont {Maeyama}}, \bibinfo {author} {\bibfnamefont {A.}~\bibnamefont {Ishizawa}}, \bibinfo {author} {\bibfnamefont {T.-H.}\ \bibnamefont {Watanabe}}, \bibinfo {author} {\bibfnamefont {N.}~\bibnamefont {Nakajima}}, \bibinfo {author} {\bibfnamefont {S.}~\bibnamefont {Tsuji-Iio}}, \ and\ \bibinfo {author} {\bibfnamefont {H.}~\bibnamefont {Tsutsui}},\ }\bibfield  {title} {\enquote {\bibinfo {title} {Numerical techniques for parallel dynamics in electromagnetic gyrokinetic vlasov simulations},}\ }\href@noop {} {\bibfield  {journal} {\bibinfo  {journal} {Comput. Phys. Commun.}\ }\textbf {\bibinfo {volume} {184}},\ \bibinfo {pages} {2462--2473} (\bibinfo {year} {2013})}\BibitemShut {NoStop}%
\bibitem [{\citenamefont {Xu}\ \emph {et~al.}(2022)\citenamefont {Xu}, \citenamefont {Peng}, \citenamefont {Hao}, \citenamefont {Chen}, \citenamefont {Li}, \citenamefont {Qu}, \citenamefont {Li}, \citenamefont {Ren}, \citenamefont {He},\ and\ \citenamefont {Li}}]{xu2022gyrokinetic}%
  \BibitemOpen
  \bibfield  {author} {\bibinfo {author} {\bibfnamefont {J.}~\bibnamefont {Xu}}, \bibinfo {author} {\bibfnamefont {X.}~\bibnamefont {Peng}}, \bibinfo {author} {\bibfnamefont {G.}~\bibnamefont {Hao}}, \bibinfo {author} {\bibfnamefont {W.}~\bibnamefont {Chen}}, \bibinfo {author} {\bibfnamefont {J.}~\bibnamefont {Li}}, \bibinfo {author} {\bibfnamefont {H.}~\bibnamefont {Qu}}, \bibinfo {author} {\bibfnamefont {J.}~\bibnamefont {Li}}, \bibinfo {author} {\bibfnamefont {G.}~\bibnamefont {Ren}}, \bibinfo {author} {\bibfnamefont {X.}~\bibnamefont {He}}, \ and\ \bibinfo {author} {\bibfnamefont {Y.}~\bibnamefont {Li}},\ }\bibfield  {title} {\enquote {\bibinfo {title} {Gyrokinetic simulations of zonal flows and ion temperature gradient turbulence in hl-2a itb plasmas},}\ }\href@noop {} {\bibfield  {journal} {\bibinfo  {journal} {Phys. Plasmas}\ }\textbf {\bibinfo {volume} {29}} (\bibinfo {year} {2022})}\BibitemShut {NoStop}%
\bibitem [{\citenamefont {Phillips}(1959)}]{phillips1959example}%
  \BibitemOpen
  \bibfield  {author} {\bibinfo {author} {\bibfnamefont {N.~A.}\ \bibnamefont {Phillips}},\ }\bibfield  {title} {\enquote {\bibinfo {title} {An example of non-linear computational instability},}\ }\href@noop {} {\bibfield  {journal} {\bibinfo  {journal} {The atmosphere and the sea in motion}\ }\textbf {\bibinfo {volume} {501}},\ \bibinfo {pages} {504} (\bibinfo {year} {1959})}\BibitemShut {NoStop}%
\bibitem [{\citenamefont {Patterson~Jr}\ and\ \citenamefont {Orszag}(1971)}]{patterson1971spectral}%
  \BibitemOpen
  \bibfield  {author} {\bibinfo {author} {\bibfnamefont {G.}~\bibnamefont {Patterson~Jr}}\ and\ \bibinfo {author} {\bibfnamefont {S.~A.}\ \bibnamefont {Orszag}},\ }\bibfield  {title} {\enquote {\bibinfo {title} {Spectral calculations of isotropic turbulence: Efficient removal of aliasing interactions},}\ }\href@noop {} {\bibfield  {journal} {\bibinfo  {journal} {Phys. Fluids}\ }\textbf {\bibinfo {volume} {14}},\ \bibinfo {pages} {2538--2541} (\bibinfo {year} {1971})}\BibitemShut {NoStop}%
\bibitem [{\citenamefont {Orszag}(1971)}]{orszag1971elimination}%
  \BibitemOpen
  \bibfield  {author} {\bibinfo {author} {\bibfnamefont {S.~A.}\ \bibnamefont {Orszag}},\ }\bibfield  {title} {\enquote {\bibinfo {title} {On the elimination of aliasing in finite-difference schemes by filtering high-wavenumber components.}}\ }\href@noop {} {\bibfield  {journal} {\bibinfo  {journal} {J. Atmos. Sci.}\ }\textbf {\bibinfo {volume} {28}},\ \bibinfo {pages} {1074--1074} (\bibinfo {year} {1971})}\BibitemShut {NoStop}%
\bibitem [{\citenamefont {Shuman}(1957)}]{shuman1957numerical}%
  \BibitemOpen
  \bibfield  {author} {\bibinfo {author} {\bibfnamefont {F.~G.}\ \bibnamefont {Shuman}},\ }\bibfield  {title} {\enquote {\bibinfo {title} {Numerical methods in weather prediction: Ii. smoothing and filtering},}\ }\href@noop {} {\bibfield  {journal} {\bibinfo  {journal} {Mon. Weather Rev.}\ }\textbf {\bibinfo {volume} {85}},\ \bibinfo {pages} {357--361} (\bibinfo {year} {1957})}\BibitemShut {NoStop}%
\bibitem [{\citenamefont {Strugarek}\ \emph {et~al.}(2013)\citenamefont {Strugarek}, \citenamefont {Sarazin}, \citenamefont {Zarzoso}, \citenamefont {Abiteboul}, \citenamefont {Brun}, \citenamefont {Cartier-Michaud}, \citenamefont {Dif-Pradalier}, \citenamefont {Garbet}, \citenamefont {Ghendrih}, \citenamefont {Grandgirard} \emph {et~al.}}]{strugarek2013unraveling}%
  \BibitemOpen
  \bibfield  {author} {\bibinfo {author} {\bibfnamefont {A.}~\bibnamefont {Strugarek}}, \bibinfo {author} {\bibfnamefont {Y.}~\bibnamefont {Sarazin}}, \bibinfo {author} {\bibfnamefont {D.}~\bibnamefont {Zarzoso}}, \bibinfo {author} {\bibfnamefont {J.}~\bibnamefont {Abiteboul}}, \bibinfo {author} {\bibfnamefont {A.}~\bibnamefont {Brun}}, \bibinfo {author} {\bibfnamefont {T.}~\bibnamefont {Cartier-Michaud}}, \bibinfo {author} {\bibfnamefont {G.}~\bibnamefont {Dif-Pradalier}}, \bibinfo {author} {\bibfnamefont {X.}~\bibnamefont {Garbet}}, \bibinfo {author} {\bibfnamefont {P.}~\bibnamefont {Ghendrih}}, \bibinfo {author} {\bibfnamefont {V.}~\bibnamefont {Grandgirard}},  \emph {et~al.},\ }\bibfield  {title} {\enquote {\bibinfo {title} {Unraveling quasiperiodic relaxations of transport barriers with gyrokinetic simulations of tokamak plasmas},}\ }\href@noop {} {\bibfield  {journal} {\bibinfo  {journal} {Phys. Rev. Lett.}\ }\textbf {\bibinfo {volume} {111}},\ \bibinfo {pages} {145001} (\bibinfo {year}
  {2013})}\BibitemShut {NoStop}%
\bibitem [{\citenamefont {Imadera}\ and\ \citenamefont {Kishimoto}(2022)}]{imadera2022itb}%
  \BibitemOpen
  \bibfield  {author} {\bibinfo {author} {\bibfnamefont {K.}~\bibnamefont {Imadera}}\ and\ \bibinfo {author} {\bibfnamefont {Y.}~\bibnamefont {Kishimoto}},\ }\bibfield  {title} {\enquote {\bibinfo {title} {Itb formation in gyrokinetic flux-driven itg/tem turbulence},}\ }\href@noop {} {\bibfield  {journal} {\bibinfo  {journal} {Plasma Phys. Control. Fusion}\ }\textbf {\bibinfo {volume} {65}},\ \bibinfo {pages} {024003} (\bibinfo {year} {2022})}\BibitemShut {NoStop}%
\bibitem [{\citenamefont {Wang}, \citenamefont {Wang},\ and\ \citenamefont {Wu}(2024)}]{wang2024self}%
  \BibitemOpen
  \bibfield  {author} {\bibinfo {author} {\bibfnamefont {S.}~\bibnamefont {Wang}}, \bibinfo {author} {\bibfnamefont {Z.}~\bibnamefont {Wang}}, \ and\ \bibinfo {author} {\bibfnamefont {T.}~\bibnamefont {Wu}},\ }\bibfield  {title} {\enquote {\bibinfo {title} {Self-organized evolution of the internal transport barrier in ion-temperature-gradient driven gyrokinetic turbulence},}\ }\href@noop {} {\bibfield  {journal} {\bibinfo  {journal} {Phys. Rev. Lett.}\ }\textbf {\bibinfo {volume} {132}},\ \bibinfo {pages} {065106} (\bibinfo {year} {2024})}\BibitemShut {NoStop}%
\bibitem [{\citenamefont {Tam}(1995)}]{tam1995computational}%
  \BibitemOpen
  \bibfield  {author} {\bibinfo {author} {\bibfnamefont {C.~K.}\ \bibnamefont {Tam}},\ }\bibfield  {title} {\enquote {\bibinfo {title} {Computational aeroacoustics-issues and methods},}\ }\href@noop {} {\bibfield  {journal} {\bibinfo  {journal} {AIAA journal}\ }\textbf {\bibinfo {volume} {33}},\ \bibinfo {pages} {1788--1796} (\bibinfo {year} {1995})}\BibitemShut {NoStop}%
\bibitem [{\citenamefont {Pirozzoli}(2006)}]{pirozzoli2006spectral}%
  \BibitemOpen
  \bibfield  {author} {\bibinfo {author} {\bibfnamefont {S.}~\bibnamefont {Pirozzoli}},\ }\bibfield  {title} {\enquote {\bibinfo {title} {On the spectral properties of shock-capturing schemes},}\ }\href@noop {} {\bibfield  {journal} {\bibinfo  {journal} {J. Comput. Phys.}\ }\textbf {\bibinfo {volume} {219}},\ \bibinfo {pages} {489--497} (\bibinfo {year} {2006})}\BibitemShut {NoStop}%
\bibitem [{\citenamefont {Shu}(2020)}]{shu2020essentially}%
  \BibitemOpen
  \bibfield  {author} {\bibinfo {author} {\bibfnamefont {C.-W.}\ \bibnamefont {Shu}},\ }\bibfield  {title} {\enquote {\bibinfo {title} {Essentially non-oscillatory and weighted essentially non-oscillatory schemes},}\ }\href@noop {} {\bibfield  {journal} {\bibinfo  {journal} {Acta Numerica}\ }\textbf {\bibinfo {volume} {29}},\ \bibinfo {pages} {701--762} (\bibinfo {year} {2020})}\BibitemShut {NoStop}%
\bibitem [{\citenamefont {Lele}(1992)}]{lele1992compact}%
  \BibitemOpen
  \bibfield  {author} {\bibinfo {author} {\bibfnamefont {S.~K.}\ \bibnamefont {Lele}},\ }\bibfield  {title} {\enquote {\bibinfo {title} {Compact finite difference schemes with spectral-like resolution},}\ }\href@noop {} {\bibfield  {journal} {\bibinfo  {journal} {J. Comput. Phys.}\ }\textbf {\bibinfo {volume} {103}},\ \bibinfo {pages} {16--42} (\bibinfo {year} {1992})}\BibitemShut {NoStop}%
\bibitem [{\citenamefont {Vasilyev}, \citenamefont {Lund},\ and\ \citenamefont {Moin}(1998)}]{vasilyev1998general}%
  \BibitemOpen
  \bibfield  {author} {\bibinfo {author} {\bibfnamefont {O.~V.}\ \bibnamefont {Vasilyev}}, \bibinfo {author} {\bibfnamefont {T.~S.}\ \bibnamefont {Lund}}, \ and\ \bibinfo {author} {\bibfnamefont {P.}~\bibnamefont {Moin}},\ }\bibfield  {title} {\enquote {\bibinfo {title} {A general class of commutative filters for les in complex geometries},}\ }\href@noop {} {\bibfield  {journal} {\bibinfo  {journal} {J. Comput. Phys.}\ }\textbf {\bibinfo {volume} {146}},\ \bibinfo {pages} {82--104} (\bibinfo {year} {1998})}\BibitemShut {NoStop}%
\bibitem [{\citenamefont {Lin}\ \emph {et~al.}(1998)\citenamefont {Lin}, \citenamefont {Hahm}, \citenamefont {Lee}, \citenamefont {Tang},\ and\ \citenamefont {White}}]{lin1998turbulent}%
  \BibitemOpen
  \bibfield  {author} {\bibinfo {author} {\bibfnamefont {Z.}~\bibnamefont {Lin}}, \bibinfo {author} {\bibfnamefont {T.~S.}\ \bibnamefont {Hahm}}, \bibinfo {author} {\bibfnamefont {W.}~\bibnamefont {Lee}}, \bibinfo {author} {\bibfnamefont {W.~M.}\ \bibnamefont {Tang}}, \ and\ \bibinfo {author} {\bibfnamefont {R.~B.}\ \bibnamefont {White}},\ }\bibfield  {title} {\enquote {\bibinfo {title} {Turbulent transport reduction by zonal flows: Massively parallel simulations},}\ }\href@noop {} {\bibfield  {journal} {\bibinfo  {journal} {Science}\ }\textbf {\bibinfo {volume} {281}},\ \bibinfo {pages} {1835--1837} (\bibinfo {year} {1998})}\BibitemShut {NoStop}%
\bibitem [{\citenamefont {Chen}\ and\ \citenamefont {Parker}(2003)}]{chen2003deltaf}%
  \BibitemOpen
  \bibfield  {author} {\bibinfo {author} {\bibfnamefont {Y.}~\bibnamefont {Chen}}\ and\ \bibinfo {author} {\bibfnamefont {S.~E.}\ \bibnamefont {Parker}},\ }\bibfield  {title} {\enquote {\bibinfo {title} {A $\delta$f particle method for gyrokinetic simulations with kinetic electrons and electromagnetic perturbations},}\ }\href@noop {} {\bibfield  {journal} {\bibinfo  {journal} {J. Comput. Phys.}\ }\textbf {\bibinfo {volume} {189}},\ \bibinfo {pages} {463--475} (\bibinfo {year} {2003})}\BibitemShut {NoStop}%
\bibitem [{\citenamefont {Idomura}, \citenamefont {Tokuda},\ and\ \citenamefont {Kishimoto}(2003)}]{idomura2003global}%
  \BibitemOpen
  \bibfield  {author} {\bibinfo {author} {\bibfnamefont {Y.}~\bibnamefont {Idomura}}, \bibinfo {author} {\bibfnamefont {S.}~\bibnamefont {Tokuda}}, \ and\ \bibinfo {author} {\bibfnamefont {Y.}~\bibnamefont {Kishimoto}},\ }\bibfield  {title} {\enquote {\bibinfo {title} {Global gyrokinetic simulation of ion temperature gradient driven turbulence in plasmas using a canonical maxwellian distribution},}\ }\href@noop {} {\bibfield  {journal} {\bibinfo  {journal} {Nucl. Fusion}\ }\textbf {\bibinfo {volume} {43}},\ \bibinfo {pages} {234} (\bibinfo {year} {2003})}\BibitemShut {NoStop}%
\bibitem [{\citenamefont {Jolliet}\ \emph {et~al.}(2007)\citenamefont {Jolliet}, \citenamefont {Bottino}, \citenamefont {Angelino}, \citenamefont {Hatzky}, \citenamefont {Tran}, \citenamefont {Mcmillan}, \citenamefont {Sauter}, \citenamefont {Appert}, \citenamefont {Idomura},\ and\ \citenamefont {Villard}}]{jolliet2007global}%
  \BibitemOpen
  \bibfield  {author} {\bibinfo {author} {\bibfnamefont {S.}~\bibnamefont {Jolliet}}, \bibinfo {author} {\bibfnamefont {A.}~\bibnamefont {Bottino}}, \bibinfo {author} {\bibfnamefont {P.}~\bibnamefont {Angelino}}, \bibinfo {author} {\bibfnamefont {R.}~\bibnamefont {Hatzky}}, \bibinfo {author} {\bibfnamefont {T.-M.}\ \bibnamefont {Tran}}, \bibinfo {author} {\bibfnamefont {B.}~\bibnamefont {Mcmillan}}, \bibinfo {author} {\bibfnamefont {O.}~\bibnamefont {Sauter}}, \bibinfo {author} {\bibfnamefont {K.}~\bibnamefont {Appert}}, \bibinfo {author} {\bibfnamefont {Y.}~\bibnamefont {Idomura}}, \ and\ \bibinfo {author} {\bibfnamefont {L.}~\bibnamefont {Villard}},\ }\bibfield  {title} {\enquote {\bibinfo {title} {A global collisionless pic code in magnetic coordinates},}\ }\href@noop {} {\bibfield  {journal} {\bibinfo  {journal} {Comput. Phys. Commun.}\ }\textbf {\bibinfo {volume} {177}},\ \bibinfo {pages} {409--425} (\bibinfo {year} {2007})}\BibitemShut {NoStop}%
\bibitem [{\citenamefont {Heikkinen}\ \emph {et~al.}(2008)\citenamefont {Heikkinen}, \citenamefont {Janhunen}, \citenamefont {Kiviniemi},\ and\ \citenamefont {Ogando}}]{heikkinen2008full}%
  \BibitemOpen
  \bibfield  {author} {\bibinfo {author} {\bibfnamefont {J.~A.}\ \bibnamefont {Heikkinen}}, \bibinfo {author} {\bibfnamefont {S.~J.}\ \bibnamefont {Janhunen}}, \bibinfo {author} {\bibfnamefont {T.~P.}\ \bibnamefont {Kiviniemi}}, \ and\ \bibinfo {author} {\bibfnamefont {F.}~\bibnamefont {Ogando}},\ }\bibfield  {title} {\enquote {\bibinfo {title} {Full f gyrokinetic method for particle simulation of tokamak transport},}\ }\href@noop {} {\bibfield  {journal} {\bibinfo  {journal} {J. Comput. Phys.}\ }\textbf {\bibinfo {volume} {227}},\ \bibinfo {pages} {5582--5609} (\bibinfo {year} {2008})}\BibitemShut {NoStop}%
\bibitem [{\citenamefont {Wang}\ \emph {et~al.}(2010)\citenamefont {Wang}, \citenamefont {Diamond}, \citenamefont {Hahm}, \citenamefont {Ethier}, \citenamefont {Rewoldt},\ and\ \citenamefont {Tang}}]{wang2010nonlinear}%
  \BibitemOpen
  \bibfield  {author} {\bibinfo {author} {\bibfnamefont {W.}~\bibnamefont {Wang}}, \bibinfo {author} {\bibfnamefont {P.}~\bibnamefont {Diamond}}, \bibinfo {author} {\bibfnamefont {T.}~\bibnamefont {Hahm}}, \bibinfo {author} {\bibfnamefont {S.}~\bibnamefont {Ethier}}, \bibinfo {author} {\bibfnamefont {G.}~\bibnamefont {Rewoldt}}, \ and\ \bibinfo {author} {\bibfnamefont {W.}~\bibnamefont {Tang}},\ }\bibfield  {title} {\enquote {\bibinfo {title} {Nonlinear flow generation by electrostatic turbulence in tokamaks},}\ }\href@noop {} {\bibfield  {journal} {\bibinfo  {journal} {Phys. Plasmas}\ }\textbf {\bibinfo {volume} {17}} (\bibinfo {year} {2010})}\BibitemShut {NoStop}%
\bibitem [{\citenamefont {Chen}\ and\ \citenamefont {Zonca}(2007)}]{chen2007nonlinear}%
  \BibitemOpen
  \bibfield  {author} {\bibinfo {author} {\bibfnamefont {L.}~\bibnamefont {Chen}}\ and\ \bibinfo {author} {\bibfnamefont {F.}~\bibnamefont {Zonca}},\ }\bibfield  {title} {\enquote {\bibinfo {title} {Nonlinear equilibria, stability and generation of zonal structures in toroidal plasmas},}\ }\href@noop {} {\bibfield  {journal} {\bibinfo  {journal} {Nucl. Fusion}\ }\textbf {\bibinfo {volume} {47}},\ \bibinfo {pages} {886} (\bibinfo {year} {2007})}\BibitemShut {NoStop}%
\bibitem [{\citenamefont {Falessi}\ and\ \citenamefont {Zonca}(2019)}]{falessi2019transport}%
  \BibitemOpen
  \bibfield  {author} {\bibinfo {author} {\bibfnamefont {M.~V.}\ \bibnamefont {Falessi}}\ and\ \bibinfo {author} {\bibfnamefont {F.}~\bibnamefont {Zonca}},\ }\bibfield  {title} {\enquote {\bibinfo {title} {Transport theory of phase space zonal structures},}\ }\href@noop {} {\bibfield  {journal} {\bibinfo  {journal} {Phys. Plasmas}\ }\textbf {\bibinfo {volume} {26}} (\bibinfo {year} {2019})}\BibitemShut {NoStop}%
\bibitem [{\citenamefont {Hasegawa}\ and\ \citenamefont {Wakatani}(1987)}]{hasegawa1987self}%
  \BibitemOpen
  \bibfield  {author} {\bibinfo {author} {\bibfnamefont {A.}~\bibnamefont {Hasegawa}}\ and\ \bibinfo {author} {\bibfnamefont {M.}~\bibnamefont {Wakatani}},\ }\bibfield  {title} {\enquote {\bibinfo {title} {Self-organization of electrostatic turbulence in a cylindrical plasma},}\ }\href@noop {} {\bibfield  {journal} {\bibinfo  {journal} {Phys. Rev. Lett.}\ }\textbf {\bibinfo {volume} {59}},\ \bibinfo {pages} {1581} (\bibinfo {year} {1987})}\BibitemShut {NoStop}%
\bibitem [{\citenamefont {Diamond}\ \emph {et~al.}(2005)\citenamefont {Diamond}, \citenamefont {Itoh}, \citenamefont {Itoh},\ and\ \citenamefont {Hahm}}]{diamond2005zonal}%
  \BibitemOpen
  \bibfield  {author} {\bibinfo {author} {\bibfnamefont {P.~H.}\ \bibnamefont {Diamond}}, \bibinfo {author} {\bibfnamefont {S.}~\bibnamefont {Itoh}}, \bibinfo {author} {\bibfnamefont {K.}~\bibnamefont {Itoh}}, \ and\ \bibinfo {author} {\bibfnamefont {T.}~\bibnamefont {Hahm}},\ }\bibfield  {title} {\enquote {\bibinfo {title} {Zonal flows in plasma—a review},}\ }\href@noop {} {\bibfield  {journal} {\bibinfo  {journal} {Plasma Phys. Control. Fusion}\ }\textbf {\bibinfo {volume} {47}},\ \bibinfo {pages} {R35} (\bibinfo {year} {2005})}\BibitemShut {NoStop}%
\bibitem [{\citenamefont {Ye}\ \emph {et~al.}(2016)\citenamefont {Ye}, \citenamefont {Xu}, \citenamefont {Xiao}, \citenamefont {Dai},\ and\ \citenamefont {Wang}}]{YeJCP16}%
  \BibitemOpen
  \bibfield  {author} {\bibinfo {author} {\bibfnamefont {L.}~\bibnamefont {Ye}}, \bibinfo {author} {\bibfnamefont {Y.}~\bibnamefont {Xu}}, \bibinfo {author} {\bibfnamefont {X.}~\bibnamefont {Xiao}}, \bibinfo {author} {\bibfnamefont {Z.}~\bibnamefont {Dai}}, \ and\ \bibinfo {author} {\bibfnamefont {S.}~\bibnamefont {Wang}},\ }\bibfield  {title} {\enquote {\bibinfo {title} {A gyrokinetic continuum code based on the numerical lie transform method},}\ }\href@noop {} {\bibfield  {journal} {\bibinfo  {journal} {J. Comput. Phys.}\ }\textbf {\bibinfo {volume} {316}},\ \bibinfo {pages} {180} (\bibinfo {year} {2016})}\BibitemShut {NoStop}%
\bibitem [{\citenamefont {Xu}\ \emph {et~al.}(2017)\citenamefont {Xu}, \citenamefont {Ye}, \citenamefont {Dai}, , \citenamefont {Xiao},\ and\ \citenamefont {Wang}}]{XuPoP17}%
  \BibitemOpen
  \bibfield  {author} {\bibinfo {author} {\bibfnamefont {Y.}~\bibnamefont {Xu}}, \bibinfo {author} {\bibfnamefont {L.}~\bibnamefont {Ye}}, \bibinfo {author} {\bibfnamefont {Z.}~\bibnamefont {Dai}}, , \bibinfo {author} {\bibfnamefont {Z.}~\bibnamefont {Xiao}}, \ and\ \bibinfo {author} {\bibfnamefont {S.}~\bibnamefont {Wang}},\ }\bibfield  {title} {\enquote {\bibinfo {title} {Nonlinear gyrokinetic simulation of ion temperature gradient turbulence based on a numerical lie-transform perturbation method},}\ }\href@noop {} {\bibfield  {journal} {\bibinfo  {journal} {Phys. Plasmas}\ }\textbf {\bibinfo {volume} {24}},\ \bibinfo {pages} {082515} (\bibinfo {year} {2017})}\BibitemShut {NoStop}%
\bibitem [{\citenamefont {Dai}\ \emph {et~al.}(2019)\citenamefont {Dai}, \citenamefont {Xu}, \citenamefont {Ye}, \citenamefont {Xiao},\ and\ \citenamefont {Wang}}]{DaiCPC19}%
  \BibitemOpen
  \bibfield  {author} {\bibinfo {author} {\bibfnamefont {Z.}~\bibnamefont {Dai}}, \bibinfo {author} {\bibfnamefont {Y.}~\bibnamefont {Xu}}, \bibinfo {author} {\bibfnamefont {L.}~\bibnamefont {Ye}}, \bibinfo {author} {\bibfnamefont {X.}~\bibnamefont {Xiao}}, \ and\ \bibinfo {author} {\bibfnamefont {S.}~\bibnamefont {Wang}},\ }\bibfield  {title} {\enquote {\bibinfo {title} {Gyrokinetic simulation of itg turbulence with toroidal geometry including the magnetic axis by using field-aligned coordinates},}\ }\href@noop {} {\bibfield  {journal} {\bibinfo  {journal} {Comput. Phys. Commun.}\ }\textbf {\bibinfo {volume} {242}},\ \bibinfo {pages} {72} (\bibinfo {year} {2019})}\BibitemShut {NoStop}%
\bibitem [{\citenamefont {Wang}(2012)}]{WangPoP12}%
  \BibitemOpen
  \bibfield  {author} {\bibinfo {author} {\bibfnamefont {S.}~\bibnamefont {Wang}},\ }\bibfield  {title} {\enquote {\bibinfo {title} {Transport formulation of the gyrokinetic turbulence},}\ }\href@noop {} {\bibfield  {journal} {\bibinfo  {journal} {Phys. Plasmas}\ }\textbf {\bibinfo {volume} {19}},\ \bibinfo {pages} {062504} (\bibinfo {year} {2012})}\BibitemShut {NoStop}%
\bibitem [{\citenamefont {Wang}(2013)}]{WangPoP13}%
  \BibitemOpen
  \bibfield  {author} {\bibinfo {author} {\bibfnamefont {S.}~\bibnamefont {Wang}},\ }\bibfield  {title} {\enquote {\bibinfo {title} {Nonlinear scattering term in the gyrokinetic vlasov equation},}\ }\href@noop {} {\bibfield  {journal} {\bibinfo  {journal} {Phys. Plasmas}\ }\textbf {\bibinfo {volume} {20}},\ \bibinfo {pages} {082312} (\bibinfo {year} {2013})}\BibitemShut {NoStop}%
\bibitem [{\citenamefont {Xiao}\ \emph {et~al.}(2017)\citenamefont {Xiao}, \citenamefont {Ye}, \citenamefont {Xu},\ and\ \citenamefont {Wang}}]{XiaoCCP17}%
  \BibitemOpen
  \bibfield  {author} {\bibinfo {author} {\bibfnamefont {X.}~\bibnamefont {Xiao}}, \bibinfo {author} {\bibfnamefont {L.}~\bibnamefont {Ye}}, \bibinfo {author} {\bibfnamefont {Y.}~\bibnamefont {Xu}}, \ and\ \bibinfo {author} {\bibfnamefont {S.}~\bibnamefont {Wang}},\ }\bibfield  {title} {\enquote {\bibinfo {title} {Application of high dimensional b-spline interpolation in solving the gyro-kinetic vlasov equation based on semi-lagrangian method},}\ }\href@noop {} {\bibfield  {journal} {\bibinfo  {journal} {Commun. Comput. Phys.}\ }\textbf {\bibinfo {volume} {22}},\ \bibinfo {pages} {789--802} (\bibinfo {year} {2017})}\BibitemShut {NoStop}%
\bibitem [{\citenamefont {Dimits}\ \emph {et~al.}(2000)\citenamefont {Dimits}, \citenamefont {Bateman}, \citenamefont {Beer}, \citenamefont {Cohen}, \citenamefont {Dorland}, \citenamefont {Hammett}, \citenamefont {Kim}, \citenamefont {Kinsey}, \citenamefont {Kotschenreuther}, \citenamefont {Kritz} \emph {et~al.}}]{dimits2000comparisons}%
  \BibitemOpen
  \bibfield  {author} {\bibinfo {author} {\bibfnamefont {A.~M.}\ \bibnamefont {Dimits}}, \bibinfo {author} {\bibfnamefont {G.}~\bibnamefont {Bateman}}, \bibinfo {author} {\bibfnamefont {M.}~\bibnamefont {Beer}}, \bibinfo {author} {\bibfnamefont {B.}~\bibnamefont {Cohen}}, \bibinfo {author} {\bibfnamefont {W.}~\bibnamefont {Dorland}}, \bibinfo {author} {\bibfnamefont {G.}~\bibnamefont {Hammett}}, \bibinfo {author} {\bibfnamefont {C.}~\bibnamefont {Kim}}, \bibinfo {author} {\bibfnamefont {J.}~\bibnamefont {Kinsey}}, \bibinfo {author} {\bibfnamefont {M.}~\bibnamefont {Kotschenreuther}}, \bibinfo {author} {\bibfnamefont {A.}~\bibnamefont {Kritz}},  \emph {et~al.},\ }\href@noop {} {\bibfield  {journal} {\bibinfo  {journal} {Phys. Plasmas}\ }\textbf {\bibinfo {volume} {7}},\ \bibinfo {pages} {969--983} (\bibinfo {year} {2000})}\BibitemShut {NoStop}%
\end{thebibliography}%

\end{document}